\documentclass[aoas,MSNbibl,nameyear,seceqn,dvips]{arximspdf}
\usepackage{graphicx}
\usepackage{mathrsfs}
\doi{10.1214/11-AOAS479}
\volume{5}
\issue{4}
\pubyear{2011}
\firstpage{2572}
\lastpage{2602}

\makeatletter

\newtheorem{theorem}{Theorem}[section]
\newtheorem{lemma}{Lemma}[section]

\newcommand{\transpose}{^\top}

\makeatother

\begin{document}
\begin{frontmatter}

\title{Sparse approximations of protein structure from noisy random
projections\thanksref{T1}}
\runtitle{Reconstruction of sparse proteins}

\thankstext{T1}{Supported in part by a European Research Council
Starting Grant Award.}

\begin{aug}
\author[A]{\fnms{Victor M.} \snm{Panaretos}\corref{}\ead[label=e1]{victor.panaretos@epfl.ch}} and
\author[A]{\fnms{Kjell} \snm{Konis}\ead[label=e2]{kjell.konis@epfl.ch}}
\runauthor{V. M. Panaretos and K. Konis}
\affiliation{Ecole Polytechnique F\'ed\'erale de Lausanne}
\address[A]{Section de Math\'ematiques\\
\'Ecole Polytechnique F\'ed\'erale de Lausanne\\
EPFL Station 8, 1015 Lausanne\\
Switzerland\\
\printead{e1}\\
\hphantom{E-mail: }\printead*{e2}} 
\end{aug}

\received{\smonth{3} \syear{2010}}
\revised{\smonth{4} \syear{2011}}

%
\begin{abstract}
Single-particle electron microscopy is a modern technique that
biophysicists employ to learn the structure of proteins. It yields data
that consist of noisy random projections of the protein structure in
random directions, with the added complication that the projection
angles cannot be observed. In order to reconstruct a three-dimensional
model, the projection directions need to be estimated by use of an
ad-hoc starting estimate of the unknown particle. In this paper we
propose a methodology that does not rely on knowledge of the projection
angles, to construct an objective data-dependent low-resolution
approximation of the unknown structure that can serve as such a
starting estimate. The approach assumes that the protein admits a
suitable sparse representation, and employs discrete
$L^1$-regularization (LASSO) as well as notions from shape theory to
tackle the peculiar challenges involved in the associated inverse
problem. We illustrate the approach by application to the
reconstruction of an E. coli protein component called the \textit{Klenow
fragment}.
\end{abstract}

%
%

\end{frontmatter}

\section{Introduction}
\begin{figure}

\includegraphics{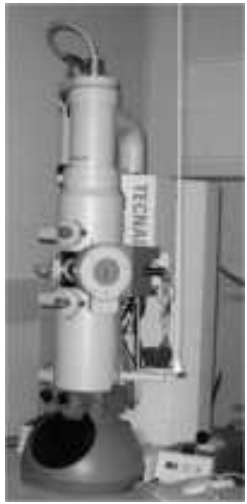}

\caption{A transmission electron microscope at the MRC Laboratory of Molecular
Biology, Cambridge, UK.}\label{figmicroscope}
\vspace*{3pt}
\end{figure}

The structure of biological macromolecules is at
the heart of the quest to understand life in purely physical terms, and
thus is fundamental to any biophysical project. A key element in
solving the structure of a protein is to be able to visualize the
protein in three dimensions, both in terms of exterior shape as well as
of interior variations. This is a challenging task given the
microscopic scale of the structures we wish to access, which can be
less than a nanometer wide. The mechanisms which enable us to gain
structural information will typically provide indirect knowledge
(posing \textit{inverse problems}), which will have to be translated into
initial structural terms in a mathematically sound way. Such mechanisms
include X-ray crystallography and electron microscopy, among others
[\citet{drenth}, \citet{glaeserbook}]. The electron microscope
(Figure \ref{figmicroscope}) in particular, is a
powerful tool that possesses important advantages over its
``competitors,'' such as high scattering power and the retainment of
phase information [\citet{chiu}, \citet{henderson}]. It
allows the retrieval of
sufficiently detailed three-dimensional representations to be able to
deduce atomic-level representations of the macromolecules of interest:
the chemical structure of the particle of interest (i.e., the amino
acid sequence) can be fitted (docked) into the density map produced via
electron microscopy to obtain the complete three-dimensional folding of
the particle [\citet{glaeserbook}].

%
%

At the most basic level, the structure of a protein is determined by
the spatial configuration of its constituent atoms. The negatively
charged electrons of each atom create a field of electric potential
surrounding the atom, and when combined, these electric potentials
create a density of potential $\rho(x,y,z)$ in space (that can be
thought of as a probability density function). This is called the \textit
{shielded Coulomb potential density}, and we usually think of a
particle as being one and the same as its potential density (it is this
density that we seek in order to dock the amino acid sequence of the
particle and completely understand its structure). When placed under an
electron beam, the potential $\rho$ causes a reduction to the beam
intensity due to electron scattering. If the beam is in the
$z$-direction, then the \textit{Abbe image formation theory}
[\citet{glaeserbook}] stipulates that the reduced intensity
recorded is
approximately given by
\[
\int_{-\infty}^{+\infty}\rho(x,y,z)\,dz+\mbox{noise}.
\]
Essentially, knowledge of the optical density recorded on the film
corresponds to knowledge of the two-dimensional marginal density of
$\rho$ in the $z$-direction, except for some minor optical effects such
as astigmatism, defocus, etc. If we were able to obtain multiple such
measurements on the same particle from various beam directions, then
determination of the three-dimensional density would amount to the
solution of a noisy tomography problem with random projection angles.
This type of problem is well understood and has been extensively
studied in the statistical literature, both methodologically
[e.g., Vardi, Shepp and Kaufman (\citeyear{vardiJASA}); Silverman et al.
(\citeyear{silveretal}); Green (\citeyear{green});
Jones and Silverman (\citeyear{jonessilver})]
and theoretically [e.g., \citet{estimationspeed},
\citet{osull2}], principally
in its positron emission tomography (PET) version.

It is impossible, however, to image the same particle under many angles
because extended exposure to the electron beam will cause chemical
bonds of the particle to break, and thus will alter the structure of
the specimen. A~means to surpass this difficulty is to crystallize many
identical particles, and thus distribute the electron dose over
multiple occurrences of the same structure, but the crystallization
process is usually cumbersome, time-consuming, and unpredictably
varying for different types of particles [\citet{glaeser}].

\textit{Single particle cryo-electron microscopy} is a technique of
electron microscopy that avoids the process of crystallization [e.g.,
\citet{glaeser}], and, as such, it is increasingly popular as a
structure determination tool in structural biology. The idea is to
image a large number of unconstrained particles in solution. The
particles rotate and diffuse freely in solution, and are then rapidly
vitrified, having assumed various different random orientations. After
preliminary processing, the data yielded are essentially a number of
noisy versions of the projected potential densities, at orientations
both \textit{random} and \textit{unknown}.

Traditional tomographic techniques break down in this setting, as these
crucially depend on the knowledge of the projection angles. In order to
be able to put these techniques to use, biophysicists attempt to
estimate the unobservable projection angles [\citet{frank}]. To this
aim, they typically assume \textit{a completely specified} low resolution
form for the unknown density. This model often relies on knowledge on
the structure of the particle gained either from other experiments or
from an ad hoc examination of the projections by eye. In some cases,
this model can be derived from the data using the so-called \textit
{projection-slice theorem} [\citet{deans}] for Radon transforms,
but the
success of this approach will depend on the level of noise in an image.
Once such a prior is given, the unknown angles are considered as
parameters to be estimated. When a set of angles is estimated, they are
used in order to obtain a traditional tomographic reconstruction
[\citet{naterrer}, \citet{deans}], and update the starting model
[\citet{frank}, \citet{glaeserbook}]. The procedure is then
iterated until it stabilizes.
Broad classes of such \textit{refinement} methods include the so-called
\textit{projection matching method} [\citet{penczek}] and \textit{3D Radon
transform method} [\citet{rademacher}]. In the first approach, the prior
model is projected over a wide range of directions, obtaining so-called
\textit{re-projections}. Each data projection is then cross-correlated
with each re-projection, and is assigned an angle corresponding to the
angle of that re-projection which produced the highest
cross-correlation. The 3D Radon approach is essentially equivalent, the
only difference being that it focuses on the Radon transform rather
than on the X-ray transform (which are of course very closely related).
Variations of these approaches also exist that try to ``integrate out''
the angles rather than estimate them: treating them as unobservable
random variables (missing data) and using an approach based on the EM
algorithm (EM here standing for Expectation--Maximization), initialized
again by some prior model for the structure of the particle
[\citet{sigworth}, \citet{bern}]. Indeed, this latter
approach draws interesting
parallels to the methodology of \citet{vardiJASA} in the case of
positron emission tomography. As already mentioned, though, what is
common to any of these strategies is that they require a completely
specified initial estimate for the structure. In cases where previous
structural information is not available, the level of noise is
relatively high, and a naked eye examination is either infeasible
(e.g., when the particle has no symmetries) or would best be avoided,
it is natural to seek approaches to obtaining ``objective'' initial
models directly from the data, in order to then initialize approaches
such as those mentioned above.

The purpose of this paper is to develop statistical tools that will
enable the construction of a data-dependent starting model in the
noisy setting encountered in practice. If the starting model is to
depend only on the data at hand, its construction will have to bypass
the unknown angles, thus requiring the approximate solution of a
tomographic problem that has a second layer of ill-posedness.
Nevertheless, it was seen in \citet{victorannals} that a~consistent
formal estimator for the \textit{shape} of the particle may be
constructed. However, the problem of the actual construction of an
estimate in a practical situation still remained open, as the formal
estimators introduced in \citet{victorannals} are only implicitly
defined. Their construction requires the solution of further inverse
problems, with severe instabilities due to the presence of noise, and
the approximate nature of the modeling framework. In this paper we
propose a framework for implementing estimators such as those proposed
in \citet{victorannals} under sparsity constraints. Our approach
combines $L^1$-regularization using Least Angles Regression with the
special geometry of the sample space to yield a procedure applicable to
actual electron microscope data. We illustrate the approach through an
artificial example and also by application to noisy single particle
projections of the so-called \textit{Klenow fragment}, a large protein
fragment that is produced during DNA polymerase interactions in E.
coli. The paper is structured as follows. Section \ref{formulation}
provides a statistical formulation of the problem, and some relevant
background. Section \ref{background} introduces the modeling framework
in which sparsity is to be understood. Our approach is presented in
Section~\ref{method+pyramid} and illustrated on an artificial sparse
density. Finally, the method is applied to single Klenow particles, and
an initial sparse approximation of the structure is obtained in Section
\ref{klenow}. Some concluding remarks are made in
Section~\ref{conclusions}.

\section{Statistical formulation}\label{formulation}

From the statistical perspective, the problem can be phrased as
follows. We wish to recover a compactly supported probability density
function $\rho(\mathbf x)$, $\mathbf x=(x_1,x_2,x_3)\transpose\in\mathbb{R}^3$,
given noisy discrete images of $N$ random projections,
\[
\breve{\rho}_n(x,y)=\int_{-\infty}^{+\infty}\rho(U_n\mathbf
x)\,dx_3,\qquad
n=1,\ldots,N,
\]
where $\{U_n\}_{n=1}^{N}$ is a collection of i.i.d. random rotation
matrices distributed according to normalized Haar measure on $\mathsf
{SO}(3)$, the group of rotations in~$\mathbb{R}^3$, that is,
\[
U_n\transpose U_n=I\qquad\mbox{a.s.},\qquad \operatorname{det}(U_n)=1\qquad
\mbox{a.s.}
\]
and
\[
WU_n\stackrel{d}{=}U_nV\qquad\forall
W,V\in\mathsf{SO}(3).
\]
The $N$ discrete noisy profile images $\{P_n\}_{n=1}^N$ are obtained by
sampling the projections $\{\breve{\rho}_n\}_{n=1}^N$ on a regular
$T\times T$ lattice, subject to corruption by additive noise,
\[
P_n(i,j)=\breve{\rho}_n(x_i,y_j)+\varepsilon_n(i,j),\qquad i,j=1,\ldots,T.
\]
It will be assumed that the noise arrays are independent, white and
Gaussian, $\varepsilon_n(i,j,)\stackrel{\mathrm{i.i.d.}}{\sim}\mathcal{N}(0,\sigma
_{\varepsilon}^2)$. The (more or less) standard problem of tomography
would be described by
%
%
\begin{equation}\label{standardproblem}
\mbox{Recover }\rho(\mathbf x)\mbox{ given }\{(P_n,U_n)\}
_{n=1}^N.
\end{equation}
However, in the single particle setup, the rotations $\{U_n\}$ are
unobservable, leading to the perturbed problem
%
%
\begin{equation}\label{randomtomography}
{\mbox{Recover }\rho(\mathbf x)\mbox{ given }\{P_n\}
_{n=1}^N}.
\end{equation}
The difference between these two problems is fundamental. Every
established technique for solving problem \ref{standardproblem} (e.g.,
based on singular value decomposition, likelihood, smoothed
backprojection and Fourier methods) crucially depends on the knowledge
of the projection directions $\{U_n\}$. In the absence of these
directions,\vadjust{\goodbreak} the estimation problem is not even well defined: it is easy
to see that the density $\rho(\mathbf x)$ is unidentifiable, since any
rotated/reflected version $\rho(Q\mathbf x)$, with $Q\transpose Q=I$, will
generate data with the same distributional properties. Intuitively,
this means that one cannot recover an exact coordinate system for the
density. Although this is conceptually obvious, it can be a serious
hurdle to statistical estimation: for example, if one wishes to
parametrize the unknown density using a Fourier expansion, the Fourier
coefficients will not be invariant to changes of the coordinate system.

Nevertheless, the \textit{shape} of the density $\rho$ can potentially be
recovered [\citet{victorannals}]. The shape of $\rho$, denoted
$[\rho
]$, encodes the totality of characteristics of $\rho$ that are
invariant with respect to the coordinate system
\[
[\rho]=\{\rho(U\mathbf x)\dvtx U\in\mathsf{O}(3)\},
\]
where we denote the group of orthogonal transformations in $\mathbb
{R}^d$ by $\mathsf{O}(d)$. Furthermore, it was seen in the same paper
that the shape of the projection $\breve{\rho}_n$, $[\breve{\rho}_n]=\{
\breve{\rho}_n(U\mathbf x)\dvtx U\in\mathsf{O}(2)\}$, constitutes a sufficient
statistic for $[\rho]$. Hence, identifiability combined with the
sufficiency principle would lead one to consider estimating $[\rho]$ on
the basis of estimators depending on the data solely through their
shape characteristics $[\breve{\rho}_1],\ldots,[\breve{\rho}_n]$.

Unfortunately, any likelihood-type approach turns out to be
completely intractable in this setup. However, the feasibility of
extracting an estimator from the random projections, without any
recourse to the angular component, suggests that one might consider
techniques that yield inefficient estimators that are nevertheless
``efficient enough'' to serve as a starting model for an iterative
procedure that estimates angles, conducts traditional tomography, and
iterates until the reconstruction stabilizes. Less formally, one can
set to obtain a \textit{rough initial approximation} that nevertheless
captures the essential features of the object that are required to
obtain a first set of angular estimates. In the next section we
formulate these approximations through a class of \textit{sparse radial
mixtures}. These provide, on the one hand, a~means to fruitfully
parametrize the notion of shape, and, on the other hand, a~natural way
to impose sparsity.

\section{Sparse approximations by radial mixtures}\label{background}

The key to our approach is the realization that approximating the
unknown density by a relatively simple object suffices, if the aim is
to obtain a starting reconstruction. Indeed, ad-hoc starting models
used by biophysicists often consist of collections of solid spheres.
The class of approximate models that we shall be pursuing is that of
\textit{radial mixtures},
%
%
\begin{eqnarray}\qquad
\rho(\mathbf x)&=&\sum_{k=1}^{K}q_k\phi(\mathbf x-\bolds\mu_k),\qquad
K\geq1,\qquad
\{\bolds\mu_k\}\subset\mathbb{R}^3,\qquad q_k>0,\nonumber\\[-8pt]\\[-8pt]
\sum_{i=1}^{K}q_k&=&1\nonumber
\end{eqnarray}
with $\phi(\cdot)$ a radial probability density function on $\mathbb
{R}^3$ (e.g., an isotropic Gaussian density), that is,
%
%
\begin{equation}\label{radialproperty}
\phi(\mathbf y)=\phi(U \mathbf y)\qquad \forall\mathbf y\in
\mathbb{R}^3, U\in\mathsf{O}(3).
\end{equation}
Radial mixtures comprise a flexible yet tractable class of models for
density estimation [see, e.g., \citet{hastietibs}, Chapter 6,
Section 7]. The
choice of
this class is especially well suited to this problem, as it offers two
technical advantages:
\begin{longlist}[(1)]
\item[(1)] Good behavior under rotation and projection: the rotated version
of~$\rho$ according to $U\in\mathsf{SO}(3)$ is given by
\[
(U\rho)(\mathbf x)=\rho(U\transpose\mathbf x)=\sum_{k=1}^{K}q_k\phi(U\transpose
\mathbf x-\bolds\mu_k)\stackrel{{\mbox{\fontsize{8.36}{11}\selectfont{(\ref{radialproperty})}}}}{=}\sum
_{k=1}^{K}q_k\phi(\mathbf x-U\bolds\mu_k),
\]
that is, by a radial mixture of the same densities with the same mixing
coefficients, but centered at the rotated location parameters $\{U\bolds\mu
_k\}$. The projected density at orientation $U$ will then be given by
\begin{eqnarray*}
\int_{-\infty}^{+\infty}(U\rho)(x_1,x_2,x_3)\,dx_3&=&\sum_{k=1}^{K}q_k\int
_{-\infty}^{+\infty}\phi(\mathbf x-U\bolds\mu_k)\,dx_3\\
&=&\sum_{k=1}^{K}q_k\varphi
\bigl(H(\mathbf x-U\bolds\mu_k)\bigr),
\end{eqnarray*}
where $H$ is the identity matrix with its last row deleted, and $\varphi
$ is the (unique) two-dimensional marginal of $\phi$,
\[
H=
\left[\matrix{1 & 0 & 0 \cr0 & 1 & 0}\right],\qquad
\varphi(x_1,x_2)=\int_{-\infty}^{+\infty}\phi(x_1,x_2,x_3)\,dx_3.
\]
\item[(2)] The possibility of a finite-dimensional parametrization of the
shapes of $\rho$ and of a projection $\breve{\rho}$ using the Gram
matrix of the original and projected location parameters, respectively,
\[
[\rho]=(\mathsf{Gram}(\{\bolds\mu_k\}),\{q_k\}),\qquad [\breve{\rho
}]=(\mathsf{Gram}(\{HU\bolds\mu_k\}),\{q_k\}),
\]
where for a collection of $K$ vectors $\{\mathbf w_j\}_{j=1}^{K}$, $\mathsf
{Gram}(\{\mathbf w_j\})$ represents the symmetric nonnegative matrix with
$(i,j)$-element equal to $\langle\mathbf w_i,\mathbf w_j\rangle$.
\end{longlist}

In Kendall's Shape Theory, Gram matrices are employed as a coordinate
system for the shape manifold induced by rigid motions
[\citet{kendallbook}, \citet{wilfle}]. Note that if the
vectors $\{\mathbf w_j\}
_{j=1}^{K}$ are arranged as the columns of a $3\times K$ matrix $\mathbf
W$, then we may simply write $\mathsf{Gram}( W)= W\transpose W$. This
Gram matrix encodes all the invariant characteristics with respect to
$\mathsf{O}(3)$ of the configuration $\{\mathbf w_j\}$, since it is
invariant under orthogonal transformations of the\vspace*{1pt} generating vectors:
for $B\in\mathsf{O}(3)$ we immediately see that\vadjust{\goodbreak}
$\mathsf{Gram}(BW)=W\transpose B\transpose BW=W\transpose W=\mathsf
{Gram}(W)$. Furthermore, given a Gram matrix of rank $p$, one can find
$K$ vectors in~$\mathbb{R}^d$, $d\geq p$, with centroid zero whose
pairwise inner
products are given by that Gram matrix (in fact, the specification of
such an ensemble amounts to merely solving nondegenerate lower
triangular linear systems of equations). Therefore, for a given density
$\phi$ (or projected density $\varphi$), the Gram matrices coupled with
the corresponding mixing proportions comprise a complete description of
the \textit{shapes} of the original and projected densities, respectively.

A further importance of this parametrization is that it provides an
interface with the finite-dimensional case, where projected shape is
better understood [Panaretos (\citeyear{aap,mathproc}),
\citet{lebarden}]. In particular, it
allows use of the following simple connection between projected shape
and original shape:
\begin{theorem}[{[\citet{victorannals}]}]\label{shapeinversion}
Let\vspace*{-2pt} $\{\mathbf w_k\}_{k=1}^{K}$ be a configuration of $K$ vectors in
$\mathbb{R}^3$ and let $U$ be a random element of $\mathsf{SO}(3)$
satisfying $WU\stackrel{d}{=}UV\stackrel{d}{=}U$, for any $W,V\in\mathsf
{SO}(3)$. Then
\[
\mathbb{E}[\mathsf{Gram}(\{HU\mathbf w_k\}_{k=1}^{K})
]=\tfrac{2}{3}\mathsf{Gram}(\{\mathbf w_k\}_{k=1}^{K}),
\]
where $H$ is the $3\times3$ identity matrix with its last row deleted.
\end{theorem}

Based on this result, \citet{victorannals} proved, for known $K$ and
under the assumption that for $i\neq j$ we have $q_i\neq q_j$, that the
hybrid maximum likelihood/method of moments estimator $\hat{\rho
}(\mathbf x)$,
%
%
\begin{equation}\label{estimator}
\hat{\rho}(\mathbf x)=\sum_{k=1}^{K}\hat{q}_k\phi
(\mathbf x -\hat{\bolds\mu}_k)
\end{equation}
is consistent modulo $\mathsf{O}(3)$, as the resolution of each image
$T\times T$ and the number of projections $N$ grows. Here, $\hat
{\bolds\mu}_k$ is any collection of $K$ vectors in~$\mathbb{R}^3$ with
Gram matrix
\[
\hat{G}=\frac{3}{2N}\sum_{n=1}^{N}\mathsf{Gram}(\{\widehat
{HU_n\bolds\mu_k}\}_{k=1}^{K}).
\]
The $\{\hat{q}_k\}$ and $\{\widehat{HU_n\bolds\mu_k}\}$ are maximum
likelihood estimators of the common mixing proportions and the
individual projected location parameters for each profile, stemming
from the loglikelihood
%
%
\begin{eqnarray}\label{loglikelihood}
&&\ell( \{{q}_k\},\{{HU_n\bolds\mu_k}\})\nonumber\\[-8pt]\\[-8pt]
&&\qquad\propto-\frac
{1}{NT^2}\sum_{n=1}^{N}\sum_{i=1}^{T}\sum_{j=1}^{T}\Biggl\{P_n(i,j)-\sum
_{k=1}^{K}q_k\varphi\bigl(H(\mathbf x-U_n\bolds\mu_k)\bigr)\Biggr\}^2.
\nonumber
\end{eqnarray}
The latter loglikelihood stems from the independence between
projections and between pixels, and the Gaussian assumption on the
noise. Notice that each of the vectors $HU_n\bolds\mu_k$ is treated as a
separate parameter.

\section{Reconstruction of a nearly black protein}\label{method+pyramid}

Although the latter development provides a consistent solution to the
problem from a theoretical perspective, it does not provide a solution
to the practical problem. The estimator defined formally as $\hat
{\rho}$ cannot be readily constructed given a data set of projections,
as it is implicitly defined through the likelihood equation~(\ref
{loglikelihood}). The optimization of the objective function given by
the latter equation is a separate challenge of its own---not in terms
of computational tractability, but in terms of accuracy and stability.
Among the reasons for this is the dimension of the search space (which
is $2KN+K$). This can to some extent be mitigated, if one is to obtain
separate likelihood estimates within each projection image, breaking
the overall problem into $N$ independent problems, each with search
space dimension $3K$ (and then seek a~global estimate for the mixing
proportions). But, more importantly, it is the highly nonlinear form
of the objective function (\ref{loglikelihood}) and the possibility of
parameters being almost unidentifiable (when projected means fall close
to one another) that presents the most serious complications in a
practical reconstruction. The objective function admits multiple local
optima that are in addition unstable to minor perturbations of the
noise term (the search surface has multiple relatively flat peaks).
This instability of the nonlinear likelihood function is a
manifestation of an inherent ill-posedness, which is clearly revealed
once we re-express the problem as a collection of $N$ deconvolution
problems to be solved given discrete data:
\begin{eqnarray*}
&&\{(\widehat{HU_n\bolds\mu_1},\ldots,\widehat{HU_n\bolds\mu
_K});(\hat{q}_1,\ldots,\hat{q}_K)\}
\\
&&\qquad :=\arg\min\Biggl\|P_n(i,j)-\varphi(\mathbf x)\ast\sum
_{k=1}^{K}q_K\delta(\mathbf x-HU_n\bolds\mu_k)\Biggr\|_2^2.
\end{eqnarray*}

Here, $\delta$ denotes Dirac's delta function.
In this format, the problem is\vspace*{1pt} seen to be a linear inverse problem in
the unknown function $h(\mathbf x):=\sum_{k=1}^{K}q_K\delta(\mathbf x-HU_n\bolds\mu
_k)$. The solution of such a problem would require regularization
through the imposition of some norm penalty on the function $h$ that we
wish to recover. This cannot work here because the unknown function to
be recovered is a Dirac comb---which is not an element of $L^2$ and
hence does not allow Hilbert space regularization methods. This is
simply a different way of saying that the problem cannot be treated as
a linear one: if we are interested in the locations themselves (the
spikes), the problem is fundamentally nonlinear.

Our basic idea to tackle this problem is to turn the drawback of
``singularity'' into an advantage by transforming the nonlinear problem
into a~linear problem through discretization of the solution search
space. While the function we seek to recover is not well behaved when
considering it as defined over a continuous domain, it reduces to a
very simple object once thought of as a high-dimensional vector. This
simplicity is reflected through \textit{sparseness}.

In particular, suppose that we relax our search problem and ask to
recover the image pixels that contain spikes, rather than the precise
spike locations themselves. Choose a projection $n=n_0$ and omit the
index for simplicity. Then, the problem can be approximately expressed
via the following linear equation:
%
%
\begin{equation}
{\mathcal{P}}_{T^2\times1}=
{\mathcal{X}}_{T^2\times T^2}
{\beta}_{T^2\times1}+{\varepsilon}_{T^2\times1}.
\end{equation}
Here, $\mathcal{P}$ is the vectorized image, obtained by stacking the
columns of the image matrix $P$. The matrix $\mathcal{X}$ is
constructed as follows: the $p$th column of~$\mathcal{X}$ is a
vectorized (by column) image (which we call a \textit{base profile})
generated by placing a single density $\varphi$ at the center of the
$j$th pixel. More precisely, let $u_j$ be the center of the $j$th
pixel. Then, the $p$th column of $\mathcal{X}$ is given by the vector
\[
\{\varphi(u_j-u_p)\}_{j=1}^{T^2},
\]
where $j$ runs so that the pixels are arranged in column-major order.
The parameter vector $\beta$ is a $T^2\times1$ vector containing at
most $K$ nonzero entries:
\[
\|\beta\|_0\leq K.
\]
These entries reveal which pixels contain spikes. Since the entire
density must be contained within the image boundaries, we a priori fix
entries of $\beta$ that correspond to pixels near the boundary to be
zero (or, alternatively, drop the corresponding columns from the matrix
$\mathcal{X}$). Finally, $\varepsilon$ is an i.i.d. Gaussian mean-zero
error vector.

Thus, in this discrete form, the problem has been reduced to a model
selection problem in linear regression: we wish to recover the nonzero
entries of $\beta$, which will reveal the approximate spike locations.
The key observation, of course, is that $\beta$ is \textit{sparse}: we
expect that $K\ll T^2$, so that we are attempting to recover a \textit
{nearly black object} in the terminology of \citet{donoho}. It
therefore seems quite appropriate to employ a shrinkage estimator in
this setting. There are various possibilities, but it is the LASSO
[\citet{lasso}] that arises as the most natural one in the setting of
this problem [see also \citet{hastietibs}]. Specifically, observe that
the nonzero entries of $\beta$ should be equal to the mixing
proportions corresponding to the respective spikes (in case multiple
spikes fall within the same pixel, then it would be the sum of the
corresponding mixing proportions). Since the object in question is a
probability density, we must have
\[
\|\beta\|_1=\sum_{i=1}^{T^2}|\beta_i|=\sum_{i=1}^{K}q_i=1,
\]
and we are naturally led to the following $L^1$-constrained least
squares problem in~$\beta$:
%
%
\begin{equation}
{\min}\|\mathcal{P}-\mathcal{X}\beta\|_2^2 \qquad\mbox{subject to } \|
\beta\|_1=1.
\end{equation}
Since the specifications of the problem determined the precise value
for the~$L_1$ penalty, there is even no need to perform
cross-validation to determine the bandwidth parameter. In practice, of
course, the total mass $m$ of the density will not be precisely known,
and may slightly differ from projection to projection. However, an
approximate value $\hat m$ can be easily estimated. Therefore, one can
employ the Least Angle Regression (LARS) algorithm [\citet{lars}] to
compute the whole LASSO path, and calibrate the results around a small
neighborhood of the bandwidth corresponding to the approximate mass
$\hat m$.

In order to illustrate the details (and effectiveness) of this discrete
regularization approach, we revisit an artificial example presented in
\citet{victorannals}, where a three-dimensional mixture of four
Gaussian kernels was to be recovered given its projections at randomly
chosen directions. The pseudo-particle potential density in three
dimensions was given by
%
%
\begin{equation}\label{eqnpyramid}
\rho(u) = \sum_{k = 1}^{4} \frac{q_{k}}{\sigma^{3} (\sqrt{2 \pi
})} \exp\biggl\{ - \frac{(u - \bolds{\mu}_{k})\transpose(u - \bolds{\mu}_{k})}{2
\sigma^{2}} \biggr\},
\end{equation}
where $u\,{=}\,(u_{x}, u_{y}, u_{z})\transpose\in\mathbb{R}^{3}$, $\sigma\,{=}\,0.46$,
$q_{1}\,{=}\,0.18$, $q_{2}\,{=}\,0.26$, $q_{3}\,{=}\,0.21$,
$q_{4}\,{=}\,0.35$, and with $\{ \bolds{\mu}_{k} \}$ given by $\bolds{\mu}_{1}\,{=}\,(0, 0.8,
-0.3)\transpose$, $\bolds{\mu}_{2}\,{=}\,(0.7, -0.4, -0.3)\transpose$,
$\bolds{\mu}_{3}\,{=}\allowbreak(-0.7, -0.4, -0.3)\transpose$, $\bolds{\mu}_{4}\,{=}\,(0, 0, 0.8)\transpose$. The
corresponding signal-to-noise level for the projections (understood as
the ratio of the signal to the noise variance) was at the level of $61:1$.

The method employed in \citet{victorannals} to perform the
deconvolutions required for the construction of the estimator was a
direct spectral approach based on results on Toeplitz forms
[\citet{grenander}, \citet{pisarenko}]. The approach
performed well on noiseless
projections, but would fail completely even with very small amounts of
noise. This effect is easily anticipated as the Toeplitz form approach
amounts to an approximate discrete version of deconvolution by
unregularized inversion of Fourier coefficients---which is bound to be highly
unstable in the presence of noise.

\begin{figure}
\begin{tabular}{@{}c@{\hspace*{3pt}}c@{\hspace*{3pt}}c@{}}

\includegraphics{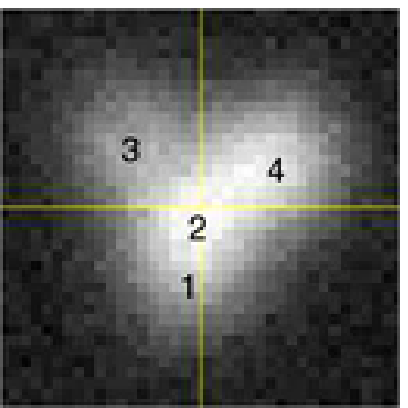}
 & \includegraphics{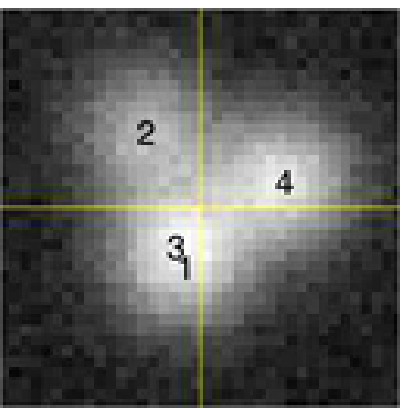} & \includegraphics{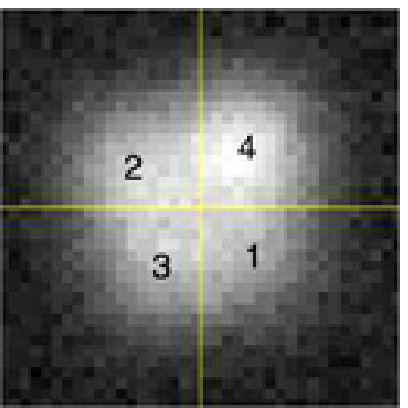}\\
(a) & (b) & (c)
\end{tabular}
\caption{Three random noisy profiles of the mixture
(\protect\ref{eqnpyramid}). The digits indicate the true locations of
the projected
component means.}\label{figprofiles}
\end{figure}

\begin{figure}[b]
\begin{tabular}{@{}cc@{}}

\includegraphics{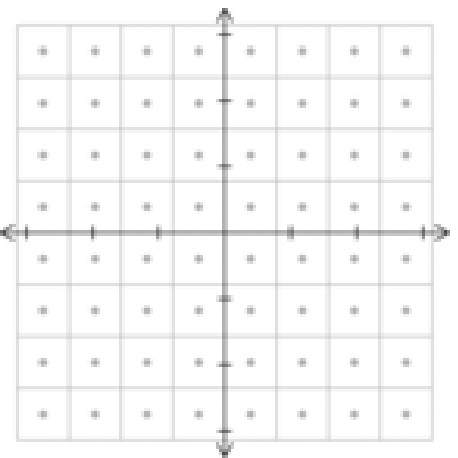}
 & \includegraphics{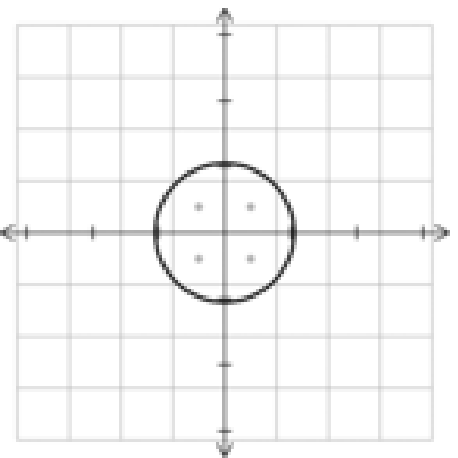} \\
(a) & (b)
\end{tabular}
\caption{Illustration of the restriction of the support for a discrete
profile. \textup{(a)} A discrete profile with $T = 8$. The pixel centers $u_{p}$
are denoted by gray dots.
\textup{(b)} The set of candidate means $\mathscr{M}$ for the convolution
matrix $\mathcal X$.}\label{figdiscprof}
\end{figure}

In order to produce a reconstruction in the presence of noise, we
implement the LASSO deconvolution approach on the basis of $N=150$
noisy random discrete profiles. The typical profile (Figure
\ref{figprofiles}) is given by a discretized image
$\mathcal{P}=\{P(i,j)\}$ defined as
%
%
\begin{equation}\label{eqndiscprof}
P(i,j) =\sum_{k = 1}^{K} q_{k} \varphi(u_{ij} | \tilde{\mu}_{k},\sigma
^2)+\varepsilon(i,j),\qquad i,j=1,\ldots,64,
\end{equation}
where $\varepsilon(i,j)$ are i.i.d. Gaussian with mean 0 and standard
deviation $10^{-4}$; $\varphi(\cdot|\nu,\sigma^2)$ is a spherical
%
%
bivariate Gaussian density with mean $\nu$ and variance~$\sigma^2$;
$u_{ij}$ is the center of the $(i,j)$th image pixel; $\{\tilde{\mu
}_{k}\}_{k = 1}^{K}$ are the locations of the $4$ (unobservable)
projected means in that profile:
\[
\tilde\mu_k:= HU\bolds\mu_k.
\]
Since each image contains a region that is known a priori to be
``empty'' (i.e., does not contain a projected mean), we limit our
interest to pixels in the complement of that region. Let $\mathscr{M}$
be the set of indices $p\in\{1,\ldots,T^2\}$ such that the pixel centers
$u_{p}$ satisfy $\|u_{p}\| < w$ (Figure \ref{figdiscprof}). We call the elements of $\{u_p\dvtx p\in
\mathscr{M}\}$ candidate means and build the convolution matrix
$\mathcal{X}$ as
\[
\mathcal{X}_{j,p} = \varphi(u_{j}| u_p, \sigma^2),\qquad
j\in\{1,\ldots,T^2\},
p\in\mathscr{M}.
\]
The choice of the tuning parameter $w$ is made so as to ensure that the
base profiles integrate to (approximately) one. In the specific
example, the choice $w = \pi/ 3$ is seen to be sufficient.

%
%

\begin{figure}
\begin{tabular}{@{}c@{\hspace*{4pt}}c@{\hspace*{4pt}}c@{}}

\includegraphics{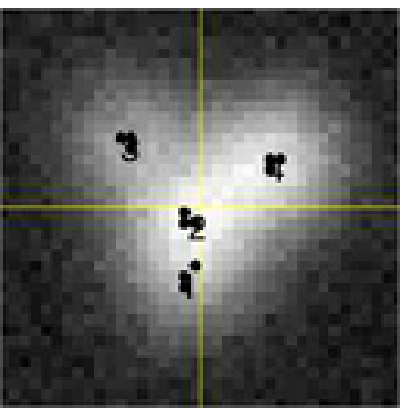}
 & \includegraphics{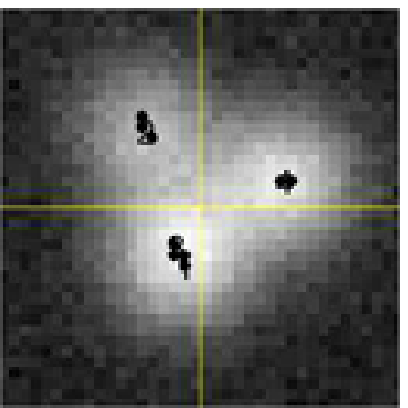} & \includegraphics{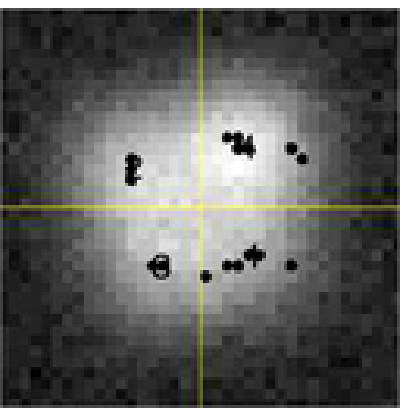}\\
(a) & (b) & (c)
\end{tabular}
\caption{The same three random profiles, with black dots indicating the
locations of the nonzero LASSO coefficient estimates. The digits
indicate the true locations of the projected component means.}\label{figld}
\end{figure}

We used the LARS algorithm (in particular, the \texttt{lars} function
in the lars package
[\citet{larsR}] for the R Project for Statistical Computing
[\citet{R}])
in order to compute the complete regularization path (in $t$) for the
LASSO problem
%
%
\begin{equation}
{\min}\|\mathcal{P}-\mathcal{X}\beta\|_2^2 \qquad\mbox{subject to } \|
\beta\|_1\leq t,
\end{equation}
and retained the parameter estimates $\hat\beta$ provided for $t$
slightly less than the estimated mass $\hat m$ (to avoid overfitting).
The latter was estimated by the average total intensity of the
projections. Figure \ref{figld} depicts three characteristic
noisy random profiles, along with the pixels the LASSO picked out as
candidate locations for mean parameters. We denote the centers of these
pixels as $\{u_p\}_{p\in\mathscr A}$, where
\[
\mathscr{A}:=\{p\in\mathscr{M}\dvtx\hat\beta_p\neq0\}.
\]
Since the true locations of the projected means will almost certainly
not be contained in $\mathcal{M}$, and since the discrete
representation of the convolution will only be approximate, a fit with
precisely the right number $K$ of nonzero parameters ($K=4$ in this
case) can rarely be achieved (i.e., $|\mathscr{A}|\neq K$). However,
the nonzero parameters found by the LASSO will tend to bracket the
locations of the projected means, as can be noticed in Figure
\ref{figld}.
It therefore suffices to use a naive clustering rule to
associate nonzero LASSO parameter estimates with projected means: if
two pixels selected by the LASSO share either an edge or a corner, then
they belong to the same cluster.

%
%

Let $\{\mathscr{C}_k\}_{k=1}^{K}$ denote the $K$ clusters comprising
$\mathscr A$, so that $\mathscr{A}=\biguplus_{k=1}^{K}\mathscr{C}_k$,
where $\biguplus$ denotes a disjoint union. Then, the estimates of the
locations of the projected means are computed by taking a weighted
average of the locations of the nonzero lasso parameter estimates in
each cluster using the parameter estimates as the relative weights,
\[
\widehat{HU\bolds\mu_k}=\frac{\sum_{p\in\mathscr{C}_k}\hat\beta_p
u_p}{\sum_{p\in\mathscr{C}_k}\beta_p}.
\]

Further, the sum of the LASSO parameter estimates in each cluster
provides an initial estimate of the mixing weight associated with that
cluster. Once all of the mixing weights have been initially estimated,
the final estimates are achieved by scaling the initial mixing weights
so that they sum to~$\hat m$, the estimated total mass of the particle,
\[
\hat q_k:=\frac{\sum_{p\in\mathscr{C}_k}\hat\beta_p}{\sum_{p\in
\mathscr{A}}\hat\beta_p}\hat m.
\]

Note here that this is the estimate of the mixing proportions stemming
from a single profile (these will later be combined to produce a global
estimate). To mitigate any bias in the mixing coefficients estimates
introduced by
choosing a constraint parameter less than $1$, one may allow the
constraint parameter to increase so long as the number of clusters remains
constant. That is, the constraint parameter is increased either until it
is equal to $1$ or until any further increase would spawn a new cluster.
Often, this results in one or more additional nonzero lasso parameter
estimates joining the current clusters. 

Once a set of estimated mixing proportions and mean locations has been
obtained for each projection, these are used in order to construct the
hybrid estimator~(\ref{estimator}). An intermediate step required is
building the estimated Gram matrices for each projection consistently.
That is, we should ensure to the extent possible that estimated
location parameters that correspond to the same three-dimensional mean
should share the same index. For this reason, within each profile, the
estimated location and mixing parameters are relabeled according to the
ascending ordering of their mixing proportions (which were assumed to
be distinct); that is, the indices are assigned so that
\[
\hat q_1< \hat q_2<\cdots<\hat q_K.
\]
Once the labels have been consistently assigned to the location
estimates within each profile, one may obtain a single profile
likelihood estimate for the mixing proportions, by solving the ordinary
least squares problem obtained when plugging the estimated location
parameters into the loglikelihood (\ref{loglikelihood}).

%
%
\begin{figure}
\begin{tabular}{@{}c@{}}

\includegraphics{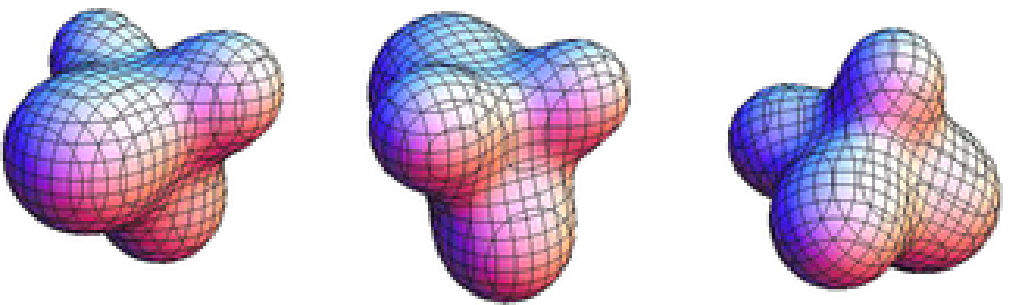}
 \\
(a)\\[4pt]

\includegraphics{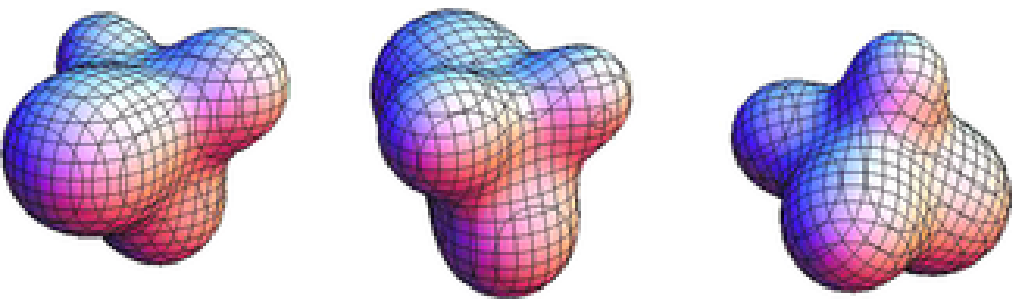}
 \\
(b)
\end{tabular}
\caption{The actual pseudo-particle density and the estimated density.
\textup{(a)} A level surface of the true pyramid density from 3
different vantage points. \textup{(b)} Corresponding level surfaces of
the estimated pyramid density.}\label{figdensitycomp}
\end{figure}

Finally, the estimated Gram matrices and the single set of estimated
mixing proportions are used to construct the hybrid estimator (\ref
{estimator}). Figure~\ref{figdensitycomp} shows the original
pseudo-particle in contrast with the estimated version. The
reconstruction was based on $53$ of the $150$ profiles in our sample,
for which four clusters were more or less clearly identifiable. In the
majority of these profiles, the $4$ clusters correspond to the
component means. However, from time-to-time, one of the clusters was a
false positive. In these cases, the smallest mixing weight was far
smaller than typical for the sample. To further filter these profiles
out, we rejected profiles with left-outlying mixing proportions (left
outlying values were omitted when calculating the mixing weights in
Table \ref{table1}).\vadjust{\goodbreak}

%
%
\begin{table}[b]
\tablewidth=290pt
\caption{Estimated mixing weights}\label{table1}
\begin{tabular*}{\tablewidth}{@{\extracolsep{\fill}}lcccc@{}}
\hline
& $\bolds{k = 1}$ & $\bolds{k = 2}$ & $\bolds{k = 3}$ & $\bolds{k = 4}$ \\
\hline
True mixing weights: $\{q_{k}\}$& 0.180 & 0.210 & 0.260 & 0.350 \\
Estimate mixing weights: $\{\hat{q}_{k}\}$ & 0.170& 0.210 & 0.263 &
0.357 \\
\hline
\end{tabular*}
\end{table}

The estimated Gram matrix and corresponding estimated location
parameters are contrasted below with the true values (the estimated
locations of components in 3 dimensions $\hat{\mu}$ can be computed by
solving a simple system of linear equations):
\begin{eqnarray*}
&G =
\pmatrix{
0.681 & -0.227 & -0.227 & -0.227 \cr
-0.227 & 0.726 & -0.254 & -0.244 \cr
-0.227 & -0.254 & 0.726 & -0.244 \cr
-0.227 & -0.244 & -0.244 & 0.716},&
\\
&\hat{G} = \pmatrix{0.696 & -0.176 & -0.279 & -0.241 \cr
-0.176 & 0.660 & -0.247 & -0.237 \cr
-0.279 & -0.247 & 0.736 & -0.209 \cr
-0.241 & -0.237 & -0.209 & 0.687},&
\\
&\pmatrix{\bolds\mu_1 & \bolds\mu_2& \bolds\mu_3 & \bolds\mu_4} =
\pmatrix{0.825 & -0.275 & -0.275 & -0.275 \cr
0.000 & 0.806 & -0.409 & -0.397 \cr
0.000 & 0.000 & 0.695 & -0.695},&
\\
&\pmatrix{\hat{\bolds\mu}_1 & \hat{\bolds\mu}_2& \hat{\bolds\mu}_3 &
\hat{\bolds\mu}_4}
= \pmatrix{0.834 & -0.211 & -0.335 & -0.289 \cr
0.000 & 0.784 & -0.405 & -0.380 \cr
0.000 & 0.000 & 0.678 & -0.678}.&
\end{eqnarray*}

\begin{figure}

\includegraphics{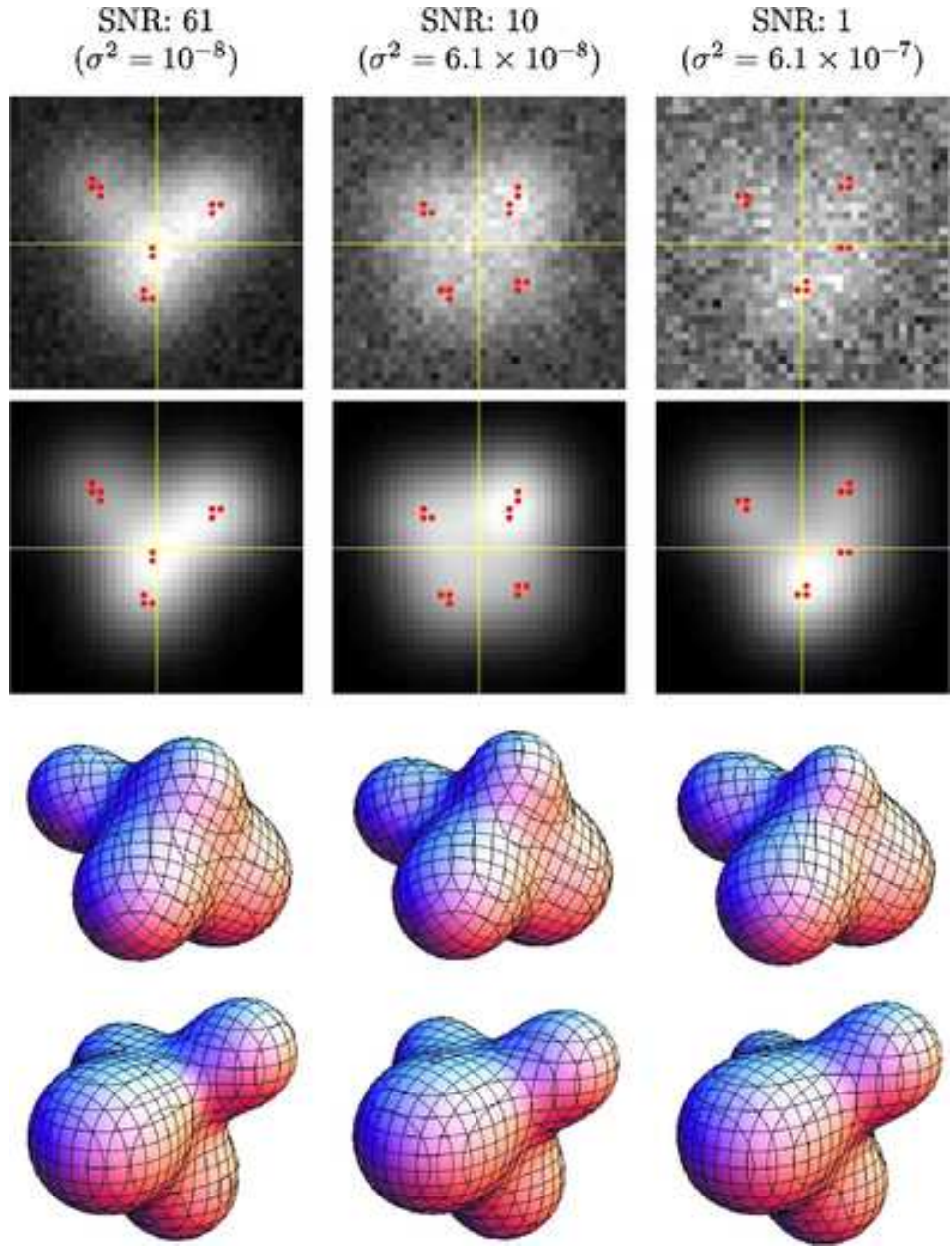}

\caption{Reconstructions under different signal-to-noise scenarios.
Each column corresponds to a different noise level. The first row
presents a typical profile along with the candidate mean positions
obtained via the LASSO. The second row presents the same profiles
without any noise, and the corresponding candidate mean positions
obtained via the LASSO. The last two rows present two different
viewpoints of the final reconstruction obtained.}\label{figsignalnoise}
\end{figure}

Interestingly enough, the reconstruction procedure was not severely
affected by higher levels of noise contamination. Even when the noise
variance was increased by two orders of magnitude, leading to a $1:1$
signal-to-noise ratio, the reconstructed version of the particle was
not significantly perturbed (see Figure \ref{figsignalnoise}). Since
%
%
it is the deconvolution step that is the most ill-posed aspect of our
approach, this can be largely attributed to a noteworthy degree of
stability exhibited by the LASSO as a means for
deconvolution.\looseness=1

\section{More on the geometry of the problem} The implementation of the
LASSO based hybrid estimator to the almost black pseudo-particle of the
previous section brings to the surface two potential issues that might
arise when implementing the procedure to actual proteins (as will be
done in Section \ref{klenow}). We consider these in the next two paragraphs.

\subsection{Using fewer projections}\label{lessprojections} The first
point relates to the usability of all the profiles available. It was
seen that several projections were not used because the viewing angles
that they represented caused problems in the construction of the
estimator. Nevertheless, the estimator constructed seemed to be rather
successful. It is therefore natural to wonder if one could do with far
fewer projections. This will become especially relevant in practical
situations where a number of projections might not present
well-identifiable mixture means. The answer is in the affirmative, that
is, one can typically use a very small number of projections, as is
illustrated by the next lemma.

\begin{lemma}\label{projectionlemma}
Let $H_1,H_2,H_3$ be projection matrices of rank 2 acting on~$\mathbb
{R}^3$ and $(\bolds\mu_1,\ldots,\bolds\mu_K)$ be an ensemble of $K$ vectors in
$\mathbb{R}^3$. If the ranges of $I-H_1,I-H_2$, $I-H_3$ are pairwise
orthogonal, then
\[
{\mathsf{Gram}}(\{\bolds\mu_k\}_{k=1}^{K})=\frac{1}{2}\sum
_{i=1}^{3}{\mathsf{Gram}}(\{H_i\bolds\mu_k\}_{k=1}^{K}).
\]
\end{lemma}
\begin{pf}
Since the rank of the projection matrices involved is 2, we may find
unit vectors $\{\mathbf e_i\}_{i=1}^{3}$ such that
\[
H_i=(I-\mathbf e_i\mathbf e_i\transpose),\qquad i=1,2,3.
\]
Furthermore, since the images of $I-H_1,I-H_2$, and $I-H_3$ are
pairwise orthogonal, it must also be that $\{\mathbf e_i\}_{i=1}^{3}$ be
pairwise orthogonal, thus constituting an orthonormal basis for $\mathbb
{R}^3$. Letting $V$ denote the $3\times K$ matrix with $(\bolds\mu
_1,\ldots,\bolds\mu_K)$ as its columns, it follows that
\begin{eqnarray*}
\sum_{i=1}^{3}{\mathsf{Gram}}(\{H_i\bolds\mu_k\}_{k=1}^{K})&=&\sum
_{i=1}^{3}V\transpose H_i\transpose H_i V=V\transpose\Biggl(\sum
_{i=1}^{3}H_i\Biggr) V\\
&=&3V\transpose V-V\transpose(\mathbf e_1\mathbf e_1\transpose+\mathbf e_2\mathbf
e_2\transpose+\mathbf e_3\mathbf e_3\transpose) V=2V\transpose V\\
&=&2{\mathsf{Gram}}(\{\bolds\mu_k\}_{k=1}^{K}).
\end{eqnarray*}
\upqed\end{pf}

Lemma \ref{projectionlemma} allows us to heuristically reinterpret the
estimator $\hat G$ of the Gram component given by
\[
\hat{G}=\frac{3}{N}\frac{1}{2}\sum_{n=1}^{N}\mathsf{Gram}(\{
\widehat{HU_n\bolds\mu_k}\}_{k=1}^{K})
\]
by thinking of it as grouping the data into $N/3$ triads of nearly
orthogonal views, forming an estimator within each triad using
Lemma \ref{projectionlemma}, and then averaging these $N/3$ estimators.

It follows that, in principle, only a few random projections at unknown
angles suffice to reconstruct a Gram matrix---provided that we can
arrange them in groups that represent views that carry information from
relatively different viewpoints. In practice, one cannot know whether
projection angles are indeed orthogonal, since they are unknown.
However, one can try to identify classes of profiles that appear to be
carrying significantly different profile information, and use these as
a proxy. The procedure is illustrated in the next section.

\subsection{Consistent construction of Gram matrices}\label{steinersection}

The second issue that became apparent from the pseudo-particle example
has to do with the consistent construction of the Gram matrices across
different profiles. This construction hinges on the assumption that the
mixing proportions are distinct. The formula defining the
pseudo-particle guaranteed that the mixing proportions were indeed
distinct, and allowed us to successfully construct the estimator. In
practice, it is natural to expect situations where the mixing
proportions for certain components are not significantly different,
leading to instabilities in the construction of the estimated Gram
matrices. To address this problem, we can take advantage of the special
geometry of the problem and, in particular, the fact that the
projections of a radial basis function cannot lie in a totally
arbitrary subspace of the set of 2D radial basis functions: the locus
of projections is highly constrained, a fact that may be exploited in
order to assign mixing proportions in a way that attempts to respect
these constraints. The constraints on the radial basis functions induce
corresponding constraints on the \textit{support} of the projected Gram
matrices, forcing this support to depend crucially on the
three-dimensional original Gram matrix (i.e., we are dealing with a
\textit{nonregular} problem). Specifically, a projected Gram matrix
cannot be any arbitrary nonnegative definite symmetric matrix. The
locus of possible projected Gram matrices $\mathscr{G}$ comprises a
smooth surface in $\mathbb{R}^{k\times k}$ of (intrinsic) dimension 2.
The idea is therefore that an arbitrary permutation of the entries of a
projected Gram matrix induces an abrupt change in its location relative
to $\mathscr{G}$, typically mapping it far from~$\mathscr{G}$. In
principle, we should thus be able to choose an arrangement of the
entries of a projected Gram matrix so as to make it ``closest'' to the
locus of ``allowed'' Gram matrices.

\begin{figure}

\includegraphics{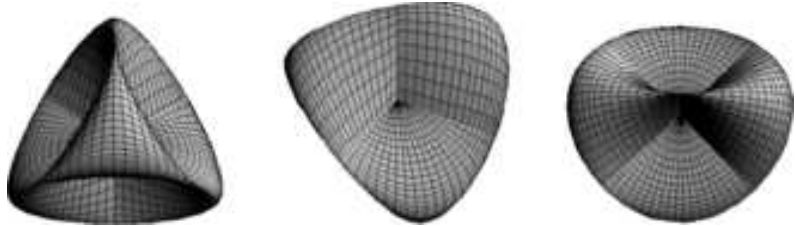}

\caption{The Roman surface from three different vantage points.}
\label{figroman}
\end{figure}

To be more precise, if $V$ is any $3\times k$ matrix such that
$G=V\transpose V$, then the projected Gram matrix at direction given by
$\mathbf e=(e_1,e_2,e_3)\transpose\in\mathbb{S}^2$ is defined as
\[
G(e)=V\transpose(I-\mathbf e \mathbf e\transpose) V.
\]
As $e$ ranges over the unit sphere, the matrix $I-\mathbf e\mathbf e\transpose$
ranges over the real projective space (the sphere with antipodal points
identified). This real projective space can be visualized in three
dimensions as the \textit{Roman surface} (see Figure \ref{figroman}),
%
%
using the mapping $\mathbb{S}^{2}\ni(e_1,e_2,e_3)\transpose\mapsto
(e_2e_3,e_1e_3,e_1e_2)\transpose$ [\citet{apery}]. The effect of
pre-multiplying by $V\transpose$ and post-multiplying by $V$ is to
stretch this Roman surface according to the singular values of $V$,
rotate it by its left singular vectors and finally shift it [much like
a~full column rank $d\times n$ matrix transforms the sphere $\mathbb
{S}^{n-1}$ into an $(n-1)$-dimensional ellipsoid in $\mathbb{R}^d$].
This can be seen directly by using Kronecker products:\looseness=1
\[
\operatorname{vec}\{G(\mathbf e)\}=\operatorname{vec}\{V\transpose(I-\mathbf e\mathbf e\transpose)
V\}=(V\transpose\otimes V\transpose)\operatorname{vec}\{(I-\mathbf e\mathbf e\transpose
)\}.
\]\looseness=0
In practice, the estimated Gram matrices will not lie precisely on the
locus~$\mathscr{G}$ since their construction is subject to error.
However, we expect that they should lie close to this surface.
Therefore, given a Gram matrix that is determined up to a permutation
of its entries, one can select the arrangement of its entries so as to
minimize its distance from the underlying locus~$\mathscr{G}$. Of
course, the exact locus $\mathscr{G}$ will not be known in practice, as
it is in bijective correspondence with the unknown three-dimensional
Gram matrix~$G$. However, an initial estimate of the surface~$\mathscr
{G}$ can be constructed using those projected Gram matrices for which
correspondences are known. The procedure is illustrated in the next
section, where we construct a sparse initial model for the potential
density of a real biological particle.

\section{Application: Sparse approximation of a Klenow fragment}\label{klenow}

We now turn to demonstrate our approach on noisy projections of an
actual biological particle called the \textit{Klenow fragment}. The
Klenow fragment is a large protein fragment that is produced in E. coli
when DNA polymerase reacts with certain enzymes [\citet{klenow}]. The
data set we will consider consists of $250$ noisy projections of the
actual known structure of the particle, produced in silico, mimicking
the behavior of the electron microscope, and kindly provided by
Professor Andres Leschziner, Harvard University [for a detailed
%
%
\begin{figure}[b]

\includegraphics{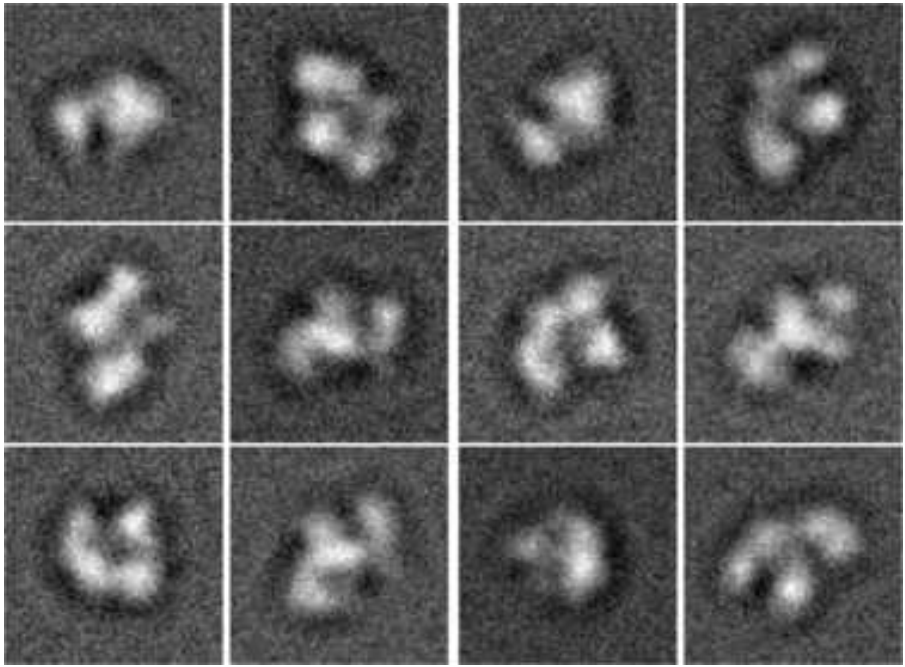}

\caption{A sample of twelve projections from the Klenow fragment data
set.}\label{figklenowdata}
\end{figure}
description of the data generation methodology, see \citet{leschziner},
Sections 2.1 and 2.2]. A sample of twelve of these projections is
depicted in Figure \ref{figklenowdata}. The projection
signal-to-noise ratio is at the level of $3:1$.

\subsection{Identifiability and blind deconvolution} A brief visual
inspection of these projections should make it immediately clear that,
unlike the synthetic particle example treated earlier, the Klenow
fragment does not fit precisely within the sparse radial mixture
framework. However, it is also apparent that if it is a coarse first
order approximation that we are interested in, then the sparse radial
model is quite reasonable. Nevertheless, the approximate nature of this
representation will have certain implications:
\begin{longlist}[(1)]
\item[(1)] The isotropic density function on which the radial representation
is based, is unknown. In essence, this means that the deconvolution
problem at hand is a blind deconvolution problem, as the point spread
function itself is poorly determined. Fortunately, we will see that the
discrete deconvolution approach based on the LASSO remains successful
even when the convolution matrix is approximate.

\item[(2)] It is likely that only a subset of the projections will be
usable, because several of the projections may involve projected means
that lie close to one another, hence pushing to the limit of unidentifiability.

\item[(3)] The mixing proportions corresponding to the best fitting radial
representation have no guarantee of being well separated. Therefore, we
will need to make use of the special geometry of the problem, as the
estimated mixing weights will not be sufficient for labeling the components.
\end{longlist}

\begin{figure}
\begin{tabular}{@{}c@{}}

\includegraphics{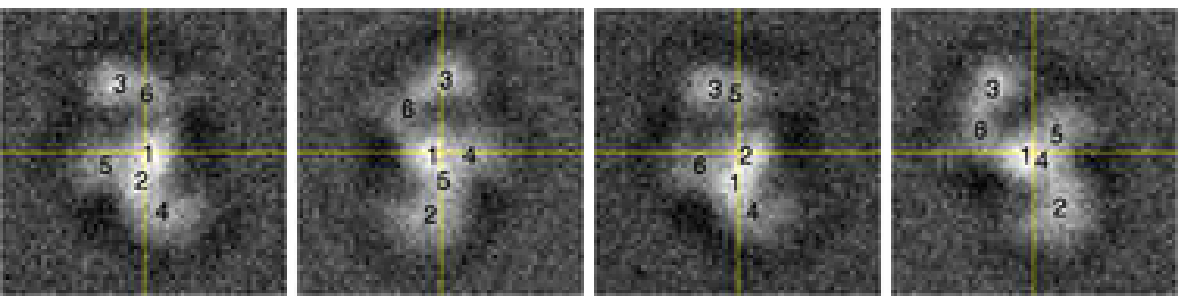}
\\
(a)\\[4pt]

\includegraphics{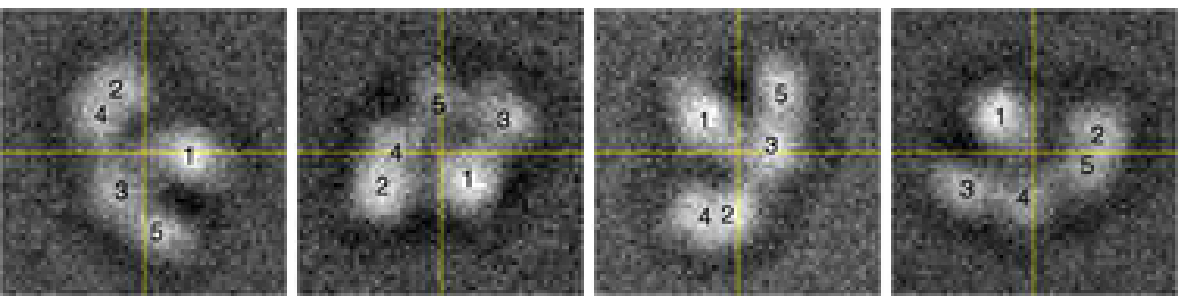}
\\
(b)\\[4pt]

\includegraphics{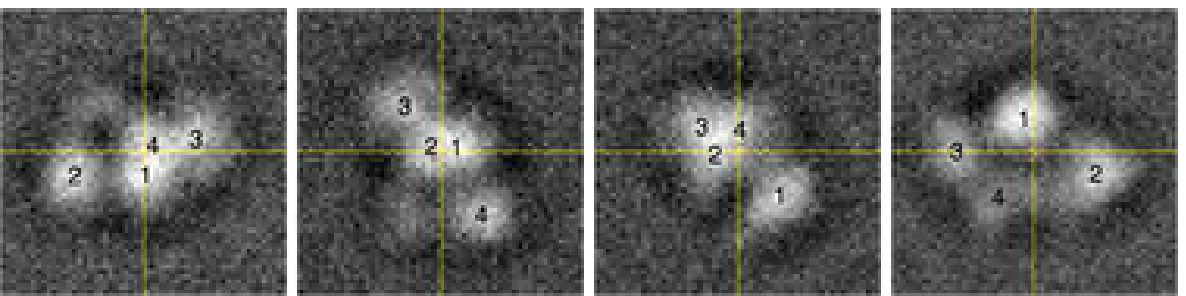}
\\
(c)
\end{tabular}
\caption{Three classes of 4 profiles. The labeling of the components
was obtained by taking the mixing weights in descending order.
\textup{(a)} Class 1: profiles with 6 identifiable projected component means.
\textup{(b)} Class 2: profiles with 5 identifiable projected component
means.
\textup{(c)}~Class~3: profiles with 4 identifiable projected component means.}
\label{figklenprofiles}
\vspace*{3pt}
\end{figure}

We begin by applying the LASSO deconvolution procedure to each of the
$250$ profiles. Since the isotropic density for the expansion is
unknown, we employ a Gaussian kernel using $\sigma= 0.224$---a value
chosen experimentally (and which will later be refined). Interestingly,
we observed that employing different kernels (or even different scale
parameters) did not significantly influence the results, provided that
$\sigma$ was not too large. Even though the point spread function was
more or less arbitrarily selected, the LASSO deconvolution procedure
produced highly sensible output (some examples are shown in Figure \ref
{figklenprofiles}), providing evidence to the effect that the
procedure is relatively robust to perturbations of the point spread
function, provided that it remains isotropic and relatively concentrated.

%
%

Since several profiles fell into the ``almost unidentifiable'' regime,
we selected three classes of profiles where the location parameters
seemed well identified and that comprised relatively different
viewpoints of the particle. The classes were constructed by choosing a
\textit{generating} profile and then selecting additional profiles that
appeared to be reflections or rotations of the generating profile.
Class $1$ consisted of profiles where the LASSO deconvolution procedure
identified 6 component means. Classes $2$ and $3$ consisted of profiles
where the LASSO deconvolution procedure identified, respectively, $5$
and $4$ component means. The three classes are shown in Figure \ref
{figklenprofiles}. Our experience showed that only very few particles
are actually required to obtain a good reconstruction and so we limited
class membership to four particles per class (in the pseudo-particle
example, we observed that as few as a~dozen could be used to produce an
excellent reconstruction).

The next steps require determining the correct labeling of the
components within each class relative to the generating profile, and
consistently combining the Gram matrix estimates from each class to
obtain an overall estimate of the Gram matrix. To describe these steps,
we use the notation~$\hat{\mu}_{k}^{(i\cdot j)}$ and~$\hat{q}_{k}^{(i\cdot j)}$
to denote, respectively, the estimated mean and mixing weight of the
$k$th component in the $j$th profile of class $i$. Additionally,
we use $\hat{\mu}^{(i\cdot j)}$ to denote the matrix with columns $\hat{\mu
}_{k}^{(i\cdot j)}$ and $\hat{q}^{(i\cdot j)}$ to denote the vector with elements
$\hat{q}_{k}^{(i\cdot j)}$, $k = 1, \ldots, K_{i}$.

\subsection{Labeling the projected component means within a class of profiles}

Since each class consists of profiles that are assumed to be
approximately rotations or reflections (plus some \textit{small}
perturbation) of the generating profile, the Gram matrix generated by
any member of the class should be \textit{close} to the Gram matrix
generated by the generating profile when the corresponding components
have the same labels. This suggests the following Procrustean algorithm
for determining the correspondences between the projected component
means in a candidate profile and those in the generating profile:

\begin{longlist}[(1)]
\item[(1)] Make a list of all possible labelings of the components in the
candidate profile.
\item[(2)] For each labeling $l$, compute the quantity $d_{l} = \|G_{R} -
G_{l}\|_{F}$ where~$G_{R}$ is the Gram matrix generated by the
reference profile, $G_{l}$ is the Gram matrix generated by the
candidate profile with labeling $l$ and \mbox{$\|\cdot\|_F$} is the Frobenius
matrix norm.
\item[(3)] Choose the labeling corresponding to the smallest $d_{l}$.
\end{longlist}

\begin{figure}
\begin{tabular}{@{}c@{\hspace*{6pt}}c@{\hspace*{6pt}}c@{}}

\includegraphics{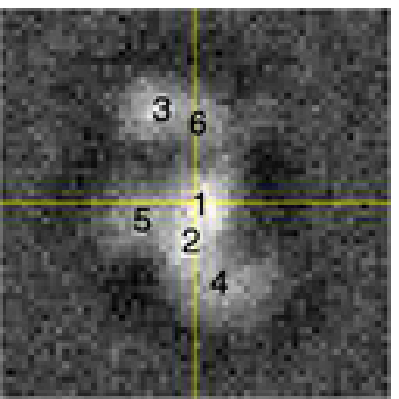}
 & \includegraphics{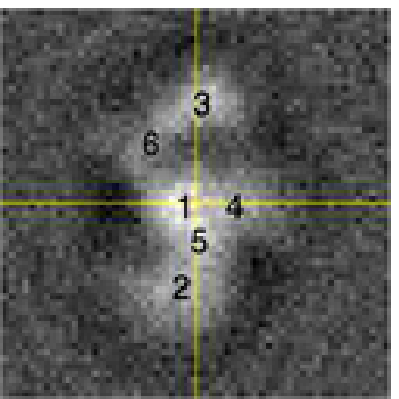} & \includegraphics{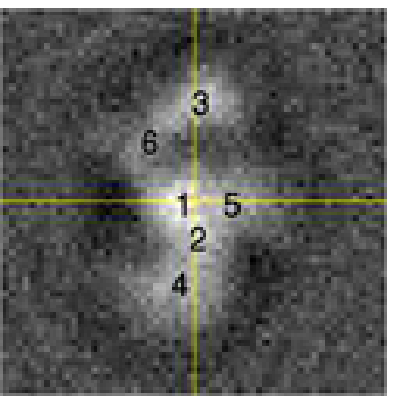}\\
(a) & (b) & (c)
\end{tabular}
\caption{The target profile is roughly a reflection of the reference
profile. The initial labeling of the components in the target profile
was obtained from the estimated mixing weights and does not agree with
the reference profile. However, the alignment algorithm finds the
correct correspondences. \textup{(a)}~Reference;
\textup{(b)} target: before; \textup{(c)} target: after.}\label{figGramOrder}
\end{figure}

An example is shown in Figure \ref{figGramOrder}. The correspondences
%
%
within each class can now be obtained by applying these steps, in turn,
to each candidate profile in the class.

\subsection{Estimating the Gram matrix}

We begin by using Theorem \ref{shapeinversion} to produce an initial
estimate of the Gram matrix for Class 1:
%
%
\begin{equation}
\tilde{G}_{1} = \frac{3}{2 \cdot4} \sum_{j = 1}^{4} \mathsf{Gram}
\bigl(\hat{\mu}^{(1\cdot j)}\bigr).
\end{equation}

It should be noted that while each individual Gram matrix within this
class encodes an ensemble that is intrinsically two-dimensional (i.e.,
has rank~2), the\vadjust{\goodbreak} Gram matrix obtained by the averaging procedure does
not necessarily encode an ensemble that can be imbedded into a
two-dimensional plane (i.e., the averaged matrix has rank higher than
2). This provides some intuition on the workings of the inversion
procedure: if all the projections within a class were identical, the
average would be exactly of rank 2, so that the averaging provides no
three-dimensional information. However, none of the class members
represent precisely the same orientation. These minor perturbations
provide some three-dimensional information, even though not dramatic:
the resulting matrix might no longer be of rank~2, but its third
singular value will be relatively small- the three-dimensional
ensemble it encodes is almost two dimensional. The intuition is that
when Gram matrices from further classes are added in (representing
significantly different orientations), the ensemble generated by the
averaged Gram matrix becomes ``more three-dimensional.''

In fact, there is no guarantee that a Gram matrix formed by averaging
several rank-2 Gram matrices will have rank 3: the rank may actually
end up being higher. For this reason, we further make a rank $3$
approximation of $\tilde{G}_{1}$ using its singular value
decomposition. Let
\[
\tilde{G}_{1} = U_{1} D_{1} V_{1}\transpose
\]
be the singular value decomposition of $\tilde{G}_{1}$ and define
\[
\hat{G}_{1} = U_{1}' D_{1}' {V_{1}'}\transpose,
\]
where $U_{1}'$ and $V_{1}'$ are, respectively, the first $3$ columns of
$U_{1}$ and $V_{1}$ and $D'$ is a diagonal matrix containing the first
$3$ singular values of $\tilde{G}_{1}$.

In class $2$, we have $K_{2} = K_{1} - 1 = 5$, hence, we assume that
one of the identified means in class $2$ has multiplicity $2$ (i.e., we
assume there are in fact~$6$ components in the true density and that
the projections of two of them fall sufficiently close in the profiles
in class $2$ that the LASSO deconvolution approach identifies them as a
single component). Also, we note that the largest component is
sufficiently distinct that it can be used reliably to identify the
first component.

%
%
\begin{figure}
\begin{tabular}{@{}cc@{}}

\includegraphics{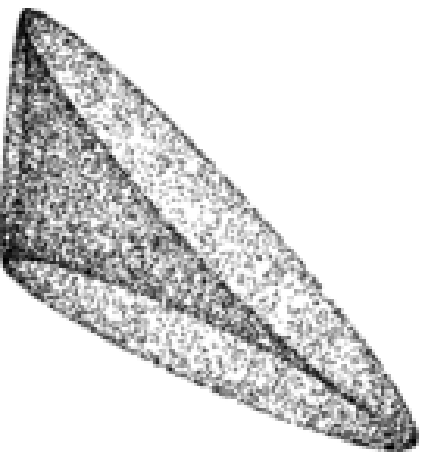}
 & \includegraphics{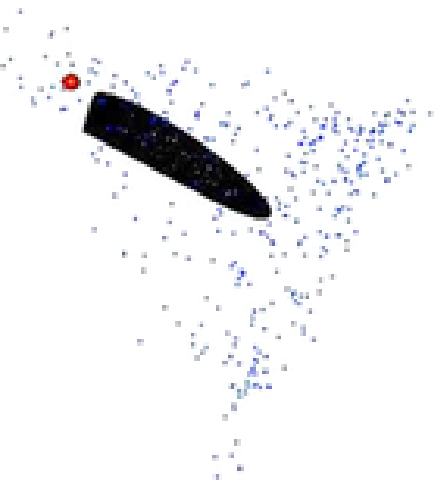} \\
(a) & (b)
\end{tabular}
\caption{\textup{(a)} Sampled points on the Roman surface generated by $\hat
{G}_1$. \textup{(b)} Point cloud of all possible permuted Gram matrices for
class 2, relative to the Roman surface generated by $\hat{G}_1$. The
red point corresponds to the point of minimum distance.}\label{fig11}
\vspace*{3pt}
\end{figure}

Following from these assumptions and in order to obtain a $6 \times6$
Gram matrix ``compatible'' with $\hat{G}_{1}$, we consider ensembles of
means of the form
\[
\bigl[\hat{\mu}^{(2\cdot j)}\ \hat{\mu}_{k}^{(2\cdot j)} \bigr]
\]
for $k = 1, \ldots, 5$ and $j = 1, \ldots, 4$. The idea is to use the
geometrical properties introduced in the previous section, to choose
that candidate Gram matrix which lies \textit{closest} to the locus of
projected Gram matrices. The latter is unknown, but we may approximate
it by the locus generated by $\hat{G}_{1}$, which constitutes itself an
estimator of the unknown three-dimensional Gram matrix. Let $\mathcal
{P} = \{P_{l}\}$ for $l \in\mathcal{L}$ be the set (with index $l$) of
all $6 \times6$ permutation matrices that leave the first row
unchanged.\vadjust{\goodbreak} We then build the set of candidate Gram matrices with elements
%
%
\begin{equation}
G_{lk} = \frac{3}{2 \cdot4} \sum_{j = 1}^{4} \mathsf{Gram} \bigl(
\bigl[\hat{\mu}^{(2\cdot j)}\ \hat{\mu}_{k}^{(2\cdot j)} \bigr] P_{l} \bigr)
\end{equation}
for all combinations of $l \in\mathcal{L}$ and $k \in\{1, \ldots, 5\}$.

Our measure of affinity to the locus generated by $\hat{G}_{1}$ is the
Euclidean distance between the candidate Gram matrix and this locus:
the stretched Roman surface generated by $\hat{G}_{1}$. In practice,
this distance is computed by randomly sampling a set of points on the
Roman surface, then taking the minimum distance between each of these
points and the candidate Gram matrix (Figure \ref{fig11}). Of course, the induced
distribution on the Roman surface will no longer be uniform, but it is
not necessary that it be. All that is required is a relatively good
coverage of the surface. We use a sample of $1\mbox{,}000$ points on the
perturbed Roman surface, defined as\looseness=1
\[
S_{n} = V_{1}' (I - u_nu_n\transpose) {V_{1}'}\transpose,\qquad n=1,\ldots,1\mbox{,}000,
\]\looseness=0
with $\{u_n\}$ being unit random vectors and $(V_{1}')\transpose
V_{1}'=\hat G_1$. We then define the distance
\[
d(G_{lk}) = {\min_{n}} \| G_{lk} - S_{n} \|_{F}.
\]
Taking $\tilde{l}$ and $\tilde{k}$ such that $d(G_{\tilde{l}\tilde
{k}})$ is minimum, we proceed to compute an initial estimate of
the Gram matrix from the profiles in classes $1$ and $2$ as
%
%
\begin{equation}
\tilde{G}_{12} = \frac{3}{2 \cdot8} \Biggl[ \sum_{j = 1}^{4}
\mathsf{Gram} \bigl( \hat{\mu}^{(1\cdot j)} \bigr) + \sum_{j = 1}^{4} \mathsf{Gram}
\bigl( \bigl[\hat{\mu}^{(2\cdot j)}\ \hat{\mu}_{\tilde{k}}^{(2\cdot j)} \bigr]
P_{\tilde{l}} \bigr) \Biggr].
\end{equation}
The final estimate $\hat{G}_{12}$ is obtained by making a rank $3$
approximation of $\tilde{G}_{12}$ using the singular value
decomposition.

Finally, we proceed analogously for the remaining class. In class $3$,
we have $K_{3} = K_{1} - 2 = 4$, hence, we assume that either one of
the identified means has multiplicity $3$, or two identified means have
multiplicity $2$. Again, to obtain a $6 \times6$ Gram matrix, we
consider ensembles of the form
\[
\bigl[ \hat{\mu}^{(3\cdot j)}\ \hat{\mu}_{k_{1}}^{(3\cdot j)}\ \hat{\mu
}_{k_{2}}^{(3\cdot j)} \bigr],
\]
where $k_{1} \leq k_{2} \in\{1, \ldots, 4\}$. The overall Gram matrix
estimate is again computed by generating a set of candidate Gram
matrices and taking the configuration that yields the smallest distance
to the stretched Roman surface generated by $\hat{G}_{12}$, then by
updating the Gram matrix estimate as above.

As a by-product of estimating the Gram matrix, we now know where to
place each of the $6$ component means in any given profile, and which
component means correspond to which from profile to profile.
Consequently, we may estimate the mixing weights and the tuning
parameter $\sigma^2$ using linear regression. Given a candidate value
for $\sigma^2$, we can construct $N$ convolution matrices $\{
\mathcal{X}_{n,\sigma^2}\}_{n=1}^N$ (corresponding to the $N$
profiles) as described in Section \ref{method+pyramid}. We thus obtain
$N$ linear regression problems, one for each projection. By\vspace*{1pt} stacking
the corresponding convolution matrices into a single $N^2\times6$
matrix, we obtain a single regression for the $6$ common mixing
weights, and estimate the latter by ordinary least squares. The
procedure can be performed for different choices of $\sigma^2$ on a
prespecified grid, retaining the set of mixing weight estimates that
correspond to the regression with the best fit. The estimate for $\sigma
^2$ thus obtained for the Klenow fragment was $\hat{\sigma}^2=0.0571$.

\begin{figure}
\begin{tabular}{@{}ccc@{}}

\includegraphics{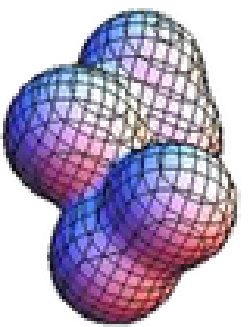}
 & \includegraphics{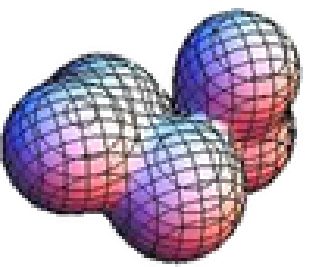} & \includegraphics{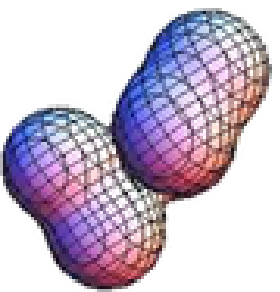}\\
(a) & (b) & (c)
\end{tabular}
\caption{Estimated density for the Klenow fragment.
\textup{(a)} View point 1; \textup{(b)} view point 2;
\textup{(c)} view point 3.}\label{figklenowrecon}
\end{figure}

\begin{figure}
\begin{tabular}{@{}cc@{}}

\includegraphics{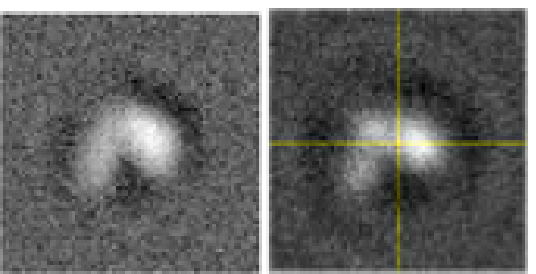}
 & \includegraphics{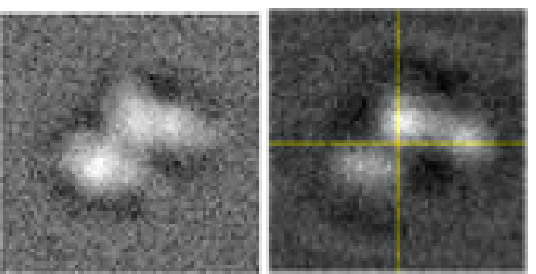} \\
(a) & (b)\\[4pt]

\includegraphics{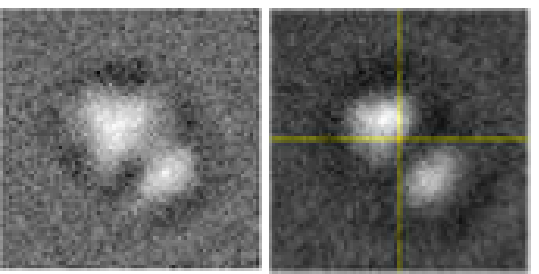}
 & \includegraphics{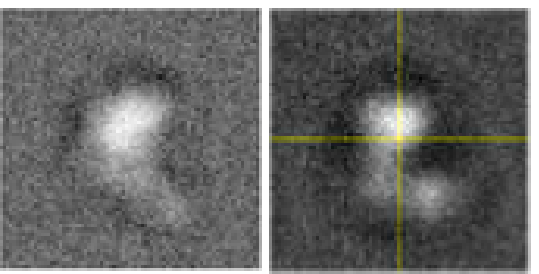} \\
(c) & (d)\\[4pt]

\includegraphics{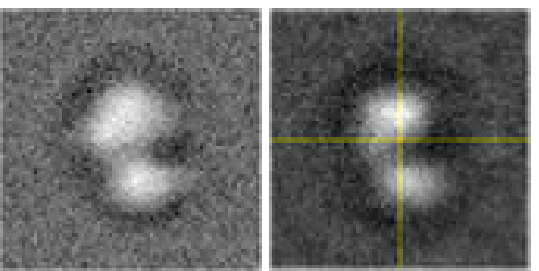}
 & \includegraphics{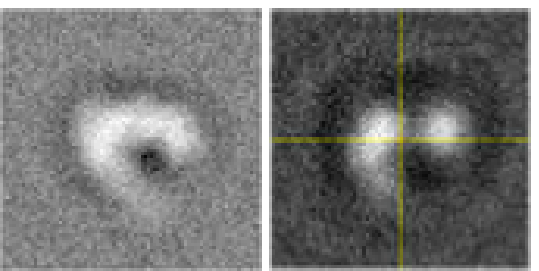} \\
(e) & (f)\\[4pt]

\includegraphics{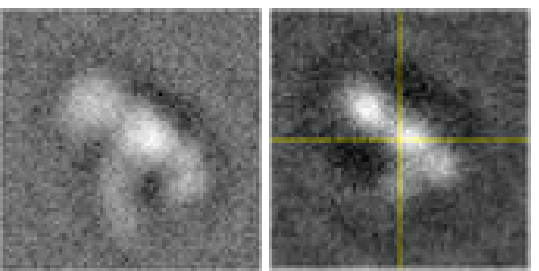}
 & \includegraphics{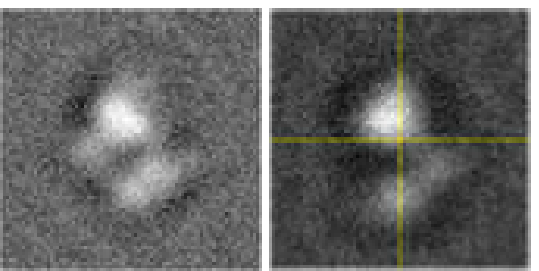} \\
(g) & (h)\\[4pt]

\includegraphics{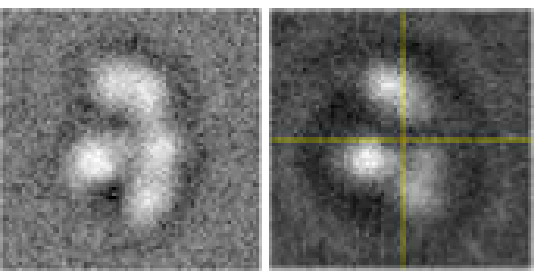}
 & \includegraphics{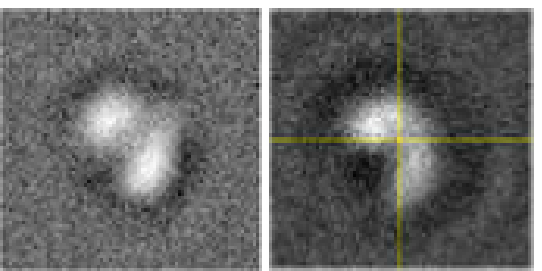} \\
(i) & (j)
\end{tabular}
\caption{Ten pairs of projections. Each pair contains an actual Klenow
fragment projection (left) coupled with a projection from our sparse
approximation (right).} \label{figcomparison}
\end{figure}

\begin{figure}
\begin{tabular}{@{}ccc@{}}

\includegraphics{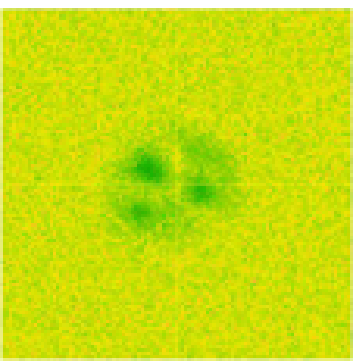}
 & \includegraphics{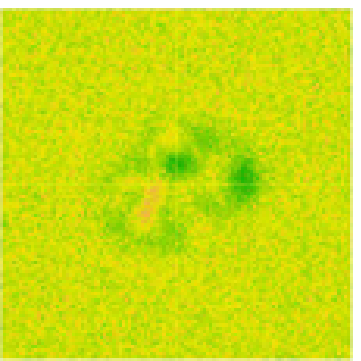} & \includegraphics{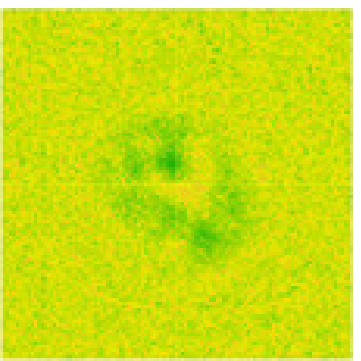}\\
(a) & (b) & (c)\\[4pt]

\includegraphics{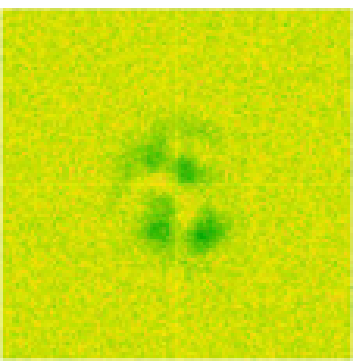}
 & \includegraphics{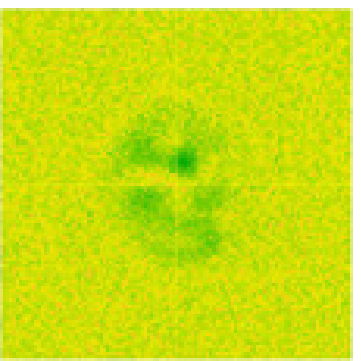} & \includegraphics{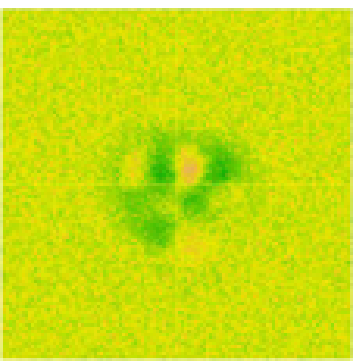}\\
(d) & (e) & (f)\\[4pt]

\includegraphics{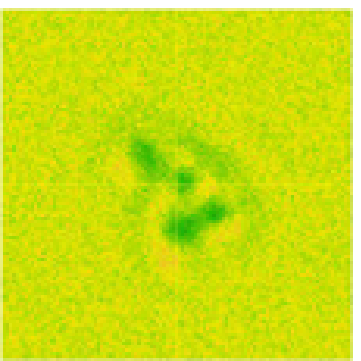}
 & \includegraphics{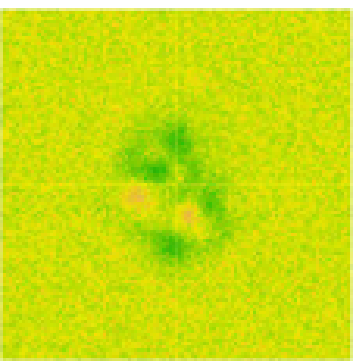} & \includegraphics{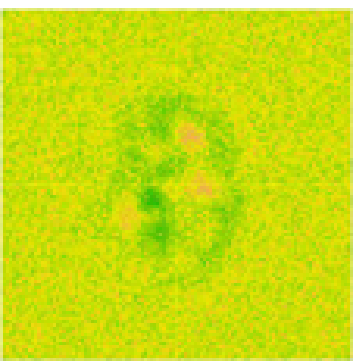}\\
(g) & (h) & (i)
\end{tabular}
\caption{Heat-maps of residual deviations of the fitted projections
from the actual projections. Each of the subfigures \textup{(a)--(i)} is
generated from the corresponding pairs \textup{(a)--(i)} in Figure
\protect\ref{figcomparison}.} \label{figresiduals}
\end{figure}

The sparse reconstruction produced by employing the estimated Gram
matrix and mixing coefficients is depicted in Figure \ref
{figklenowrecon}. In order to appreciate the ``fit'' of the sparse
%
%
reconstruction to the data, we construct noisy projections from the
reconstruction and contrast them to several typical projections of the
Klenow fragment. The projections of the reconstructed model are
constructed so as to mimic the effects that the microscope induces on
the profiles (astigmatism, defocus, contrast transfer function effects)
and so as to be characterized by a signal-to-noise ratio similar to
that of the actual projections. A sample of such contrasts is given in
Figure \ref{figcomparison}. We observe that, even though rather
%
%
sparse, the reconstructed density is able to capture the main features
of the projections quite successfully. This hints that the
reconstructed density can be appropriate to use as a starting model.
What is especially important is that the data produced by our
reconstruction seems to be highly consistent with the actual Klenow
data even for viewing angles that were \textit{not used in the
reconstruction}; in fact, this remains true even for viewing angles
that fall in the unidentifiability regime. Figure \ref{figresiduals}
%
%
shows the corresponding residual deviation heat-maps. We observe that
underestimation (corresponding to yellow/orange regions) occurs in the
regions between the components of the Gaussian mixture---evidently,
there is mass there that cannot be captured by the Gaussian mixture.
There are also some regions where overestimation occurs (darker green
regions), mostly close to the center of blob-like components of the
particle profiles, principally due to the fact that the mixing
components of the Gaussian mixture will obviously have relatively
different higher-order concentration characteristics from the blob-like
components of the actual particle. For example, in Figure \ref
{figresiduals}(f), we observe slight overestimation of the density at
locations corresponding to the center of the blob-like components of
the actual profile. These two are the only systematic patterns that
appear in the residuals, and are evidently attributed to the bias
introduced from our regularization via the Gaussian mixture model employed.

\section{Concluding remarks}\label{conclusions}

Despite the severely ill-posed nature of noisy random tomography, this
paper demonstrates that it is practically feasible to obtain useful
three-dimensional structural information on a protein given only noisy
projections at random and unknown angles. The approach proposed to this
effect rests on two basic elements: the imposition of a certain degree
of sparsity on the required reconstruction, and the exploitation of the
special geometry that is intrinsic to tomography data and provides
valuable information. Though the sparsity assumption will typically
lead to a~relatively coarse-grained approximation to the protein under
investigation, this is precisely what is required: a low-resolution
starting model that can be used as a reference structure to iteratively
recover the unknown angles to then produce a high-resolution
reconstruction based on traditional nonparametric tomographic
techniques (once the projection angles have been estimated, it is no
longer necessary to maintain the mixture model).

While it had previously been theoretically demonstrated in
\citet{victorannals} that it is feasible to reconstruct a three-dimensional
object in this setting (up to an orthogonal transformation), obtaining
an explicit reconstruction in practice remained elusive. By employing a
radial basis representation of the unknown protein, the problem of
structure determination was reduced to the problem of recovering the
Euclidean shape of the ensemble of location parameters of the radial
functions and the associated mixing coefficients. This was done in two
steps: nonlinear deconvolution and shape averaging. In the
deconvolution step, the projected location parameters had to be
identified within the noisy projections. Since the nature of the radial
expansion representation is approximate, the deconvolution problem was blind.

To tackle this problem, our approach relaxed the nonlinear
deconvolution problem into a linear problem by considering its
discretized version with an approximately chosen point spread function.
When seen in this setting, the problem falls precisely in the framework
of sparse model selection. Since the object to be recovered is a
density function, the LASSO arose as the most natural technique to
attack the problem, with an $L^1$ penalty corresponding to a
requirement on the total mass of the density to be recovered. Despite
the fact that the exact point spread function was unknown, it was seen
that the LASSO performed extremely satisfactorily where other solution
approaches break down. This was true both in the setting of artificial
examples, as well as in the setting of protein data. Once the
projections of the location parameters had been deconvolved, the
averaging step was carried out. This required the recovery of the
correspondences between location parameters in different projections.
To this effect, our approach exploited the nonregularity of the
tomography problem: it was seen that the Gram matrices of the $k\times
k$ projected components are constrained to lie in a smooth
two-dimensional subset of $\mathbb{R}^{k\times k}$, which was
identified as a deformed Roman surface. This was then exploited in
order to choose consistent correspondences across
projections.\looseness=1

The methodology was applied with success both to projection data
arising from an artificial example, as well as to projections of an
actual protein component, the Klenow fragment. Especially in the latter
case, it was seen that the sparse reconstruction recovered from the
noisy projection data can very well serve as a starting model, since
its typical projections are highly similar with those of the projection
of the true structure (Figure \ref{figcomparison}). It is therefore
likely that our approach will provide a useful means to obtaining
objective data-dependent starting models in the context of single
particle electron microscopy. From the purely statistical perspective,
the use of the LASSO in the setting of double blind deconvolution can
be of independent interest when seen in the context of estimation of
mixtures of scale-location densities of an unknown family.

\section*{Acknowledgments}
We wish to thank Professor R. M. Glaeser for many useful interactions on
the problem of initial model determination. Our thanks also go to Dr.
Richard Hall for his kind help with electron microscopy software, and
to Professor Andres Leschziner for sharing the Klenow fragment data
set. We are also thankful to two anonymous referees for their
constructive comments and suggestions.


%
\printaddresses


\begin{thebibliography}{99}

%
%
\bibitem[\protect\citeauthoryear{Apery}{1987}]{apery}
\begin{bbook}[author]
\bauthor{\bsnm{Apery},~\bfnm{F.}\binits{F.}}
(\byear{1987}).
\btitle{Models of the Real Projective Plane: Computer Graphics of
Steiner and
Boy Surfaces}.
\bpublisher{Vieweg}, \baddress{Braunschweig}.
\end{bbook}
\MR{0986729}
\endbibitem

%
%
\bibitem[\protect\citeauthoryear{Bern, Chen and Wong}{2005}]{bern}
\begin{bincollection}[author]
\bauthor{\bsnm{Bern},~\bfnm{Marshall}\binits{M.}},
\bauthor{\bsnm{Chen},~\bfnm{Jindong}\binits{J.}} \AND
\bauthor{\bsnm{Wong},~\bfnm{Hao}\binits{H.}}
(\byear{2005}).
\btitle{Avoiding local optima in single particle reconstruction}.
In \bbooktitle{Research in Computational Molecular Biology}
(\beditor{\bfnm{Satoru}\binits{S.}~\bsnm{Miyano}},
\beditor{\bfnm{Jill}\binits{J.}~\bsnm{Mesirov}},
\beditor{\bfnm{Simon}\binits{S.}~\bsnm{Kasif}},
\beditor{\bfnm{Sorin}\binits{S.}~\bsnm{Istrail}},
\beditor{\bfnm{Pavel}\binits{P.}~\bsnm{Pevzner}} \AND
\beditor{\bfnm{Michael}\binits{M.}~\bsnm{Waterman}}, eds.).
\bseries{Lecture Notes in Computer Science}
\bvolume{3500}
\bpages{118--132}.
\bpublisher{Springer}, \baddress{Berlin}.
\end{bincollection}
\MR{2304663}
\endbibitem

%
%
\bibitem[\protect\citeauthoryear{Chiu}{{1993}}]{chiu}
\begin{barticle}[author]
\bauthor{\bsnm{Chiu},~\bfnm{W.}\binits{W.}}
(\byear{{1993}}).
\btitle{What does electron cryomicroscopy provide that X-ray crystallography
and NMR spectroscopy cannot?}
\bjournal{Annu. Rev. Bioph. Biom.}
\bvolume{22}
\bpages{233--255}.
\end{barticle}
\endbibitem

%
%
\bibitem[\protect\citeauthoryear{Deans}{1993}]{deans}
\begin{bbook}[author]
\bauthor{\bsnm{Deans},~\bfnm{S.~R.}\binits{S.~R.}}
(\byear{1993}).
\btitle{The Radon Transform and Some of Its Applications}.
\bpublisher{Krieger}, \baddress{Malabar, FL}.
\end{bbook}
\MR{1274701}
\endbibitem

%
%
\bibitem[\protect\citeauthoryear{Donoho et~al.}{1992}]{donoho}
\begin{barticle}[author]
\bauthor{\bsnm{Donoho},~\bfnm{D.~L.}\binits{D.~L.}},
\bauthor{\bsnm{Johnstone},~\bfnm{I.~M.}\binits{I.~M.}},
\bauthor{\bsnm{Hoch},~\bfnm{J.~C.}\binits{J.~C.}} \AND
\bauthor{\bsnm{Stern},~\bfnm{A.~S.}\binits{A.~S.}}
(\byear{1992}).
\btitle{Maximum entropy and the nearly black object}.
\bjournal{J. Roy. Statist. Soc. Ser. B}
\bvolume{54}
\bpages{41--81}.
\end{barticle}
\MR{1157714}
\endbibitem

%
%
\bibitem[\protect\citeauthoryear{Drenth}{1999}]{drenth}
\begin{bbook}[author]
\bauthor{\bsnm{Drenth},~\bfnm{J.}\binits{J.}}
(\byear{1999}).
\btitle{Principles of Protein X-Ray Crystallography}.
\bpublisher{Springer}, \baddress{New York}.
\end{bbook}
\endbibitem

%
%
\bibitem[\protect\citeauthoryear{Efron et~al.}{2004}]{lars}
\begin{barticle}[author]
\bauthor{\bsnm{Efron},~\bfnm{B.}\binits{B.}},
\bauthor{\bsnm{Hastie},~\bfnm{T.}\binits{T.}},
\bauthor{\bsnm{Johnstone},~\bfnm{I.}\binits{I.}} \AND
\bauthor{\bsnm{Tibshirani},~\bfnm{R.}\binits{R.}}
(\byear{2004}).
\btitle{Least angle regression}.
\bjournal{Ann. Statist.}
\bvolume{32}
\bpages{407--451}.
\end{barticle}
\MR{2060166}
\endbibitem

%
%
\bibitem[\protect\citeauthoryear{Frank}{1999}]{frank}
\begin{bbook}[author]
\bauthor{\bsnm{Frank},~\bfnm{J.}\binits{J.}}
(\byear{1999}).
\btitle{Three-Dimensional Electron Microscopy of Macromolecular Assemblies}.
\bpublisher{Academic Press}, \baddress{San Diego}.
\end{bbook}
\endbibitem

%
%
\bibitem[\protect\citeauthoryear{Glaeser}{1999}]{glaeser}
\begin{barticle}[author]
\bauthor{\bsnm{Glaeser},~\bfnm{R.~M.}\binits{R.~M.}}
(\byear{1999}).
\btitle{{R}eview: {E}lectron crystallography: {P}resent excitement, a
nod to
the past, anticipating the future}.
\bjournal{J. Struct. Biol.}
\bvolume{128}
\bpages{3--14}.
\end{barticle}
\endbibitem

%
%
\bibitem[\protect\citeauthoryear{Glaeser et~al.}{2007}]{glaeserbook}
\begin{bbook}[author]
\bauthor{\bsnm{Glaeser},~\bfnm{R.~M.}\binits{R.~M.}},
\bauthor{\bsnm{Chiu},~\bfnm{W.}\binits{W.}},
\bauthor{\bsnm{Frank},~\bfnm{J.}\binits{J.}},
\bauthor{\bsnm{DeRosier},~\bfnm{D.}\binits{D.}},
\bauthor{\bsnm{Baumeister},~\bfnm{W.}\binits{W.}} \AND
\bauthor{\bsnm{Downing},~\bfnm{K.}\binits{K.}}
(\byear{2007}).
\btitle{Electron Crystallography of Biological Macromolecules}.
\bpublisher{Oxford Univ. Press}, \baddress{Oxford}.
\end{bbook}
\endbibitem

%
%
\bibitem[\protect\citeauthoryear{Green}{1990}]{green}
\begin{barticle}[author]
\bauthor{\bsnm{Green},~\bfnm{P.~J.}\binits{P.~J.}}
(\byear{1990}).
\btitle{Bayesian reconstructions from emission tomography data using a~modified
EM algorithm}.
\bjournal{IEEE T. Med. Imaging}
\bvolume{9}
\bpages{84--93}.
\end{barticle}
\endbibitem

%
%
\bibitem[\protect\citeauthoryear{Grenander and Szeg\"o}{1958}]{grenander}
\begin{bbook}[author]
\bauthor{\bsnm{Grenander},~\bfnm{U.}\binits{U.}} \AND
\bauthor{\bsnm{Szeg\"o},~\bfnm{G.}\binits{G.}}
(\byear{1958}).
\btitle{Toeplitz Forms and Their Applications}.
\bpublisher{Univ. California Press}, \baddress{Berkeley}.
\end{bbook}
\MR{0094840}
\endbibitem

%
%
\bibitem[\protect\citeauthoryear{Hastie and Efron}{2011}]{larsR}
\begin{bmisc}[author]
\bauthor{\bsnm{Hastie},~\bfnm{Trevor}\binits{T.}} \AND
\bauthor{\bsnm{Efron},~\bfnm{Brad}\binits{B.}}
(\byear{2011}).
\bhowpublished{lars: Least angle regression, Lasso and forward
stagewise. R~package version 0.9-8.}
\end{bmisc}
\endbibitem

%
%
\bibitem[\protect\citeauthoryear{Hastie, Tibshirani and
Friedman}{2001}]{hastietibs}
\begin{bbook}[author]
\bauthor{\bsnm{Hastie},~\bfnm{T.}\binits{T.}},
\bauthor{\bsnm{Tibshirani},~\bfnm{R.}\binits{R.}} \AND
\bauthor{\bsnm{Friedman},~\bfnm{J.}\binits{J.}}
(\byear{2001}).
\btitle{The Elements of Statistical Learning: Data Mining, Inference, and Prediction}.
\bpublisher{Springer}, \baddress{New York}.
\end{bbook}
\MR{1851606}
\endbibitem

%
%
\bibitem[\protect\citeauthoryear{Henderson}{2004}]{henderson}
\begin{barticle}[author]
\bauthor{\bsnm{Henderson},~\bfnm{R.}\binits{R.}}
(\byear{2004}).
\btitle{Realizing the potential of electron cryo-microscopy}.
\bjournal{Q. Rev. Biophys.}
\bvolume{37}
\bpages{3--13}.
\end{barticle}
\endbibitem

%
%
\bibitem[\protect\citeauthoryear{Johnstone and
Silverman}{1990}]{estimationspeed}
\begin{barticle}[author]
\bauthor{\bsnm{Johnstone},~\bfnm{I.~M.}\binits{I.~M.}} \AND
\bauthor{\bsnm{Silverman},~\bfnm{B.~W.}\binits{B.~W.}}
(\byear{1990}).
\btitle{Speed of estimation in positron emission tomography and related inverse
problems}.
\bjournal{Ann. Statist.}
\bvolume{18}
\bpages{251--280}.
\end{barticle}
\MR{1041393}
\endbibitem

%
%
\bibitem[\protect\citeauthoryear{Jones and Silverman}{1989}]{jonessilver}
\begin{barticle}[author]
\bauthor{\bsnm{Jones},~\bfnm{M.~C.}\binits{M.~C.}} \AND
\bauthor{\bsnm{Silverman},~\bfnm{B.~W.}\binits{B.~W.}}
(\byear{1989}).
\btitle{An orthogonal series density estimation approach to reconstructing
positron emission tomography images}.
\bjournal{J. Appl. Stat.}
\bvolume{16}
\bpages{177--191}.
\end{barticle}
\endbibitem

%
%
\bibitem[\protect\citeauthoryear{Kendall and Le}{2009}]{wilfle}
\begin{bincollection}[author]
\bauthor{\bsnm{Kendall},~\bfnm{W.~S.}\binits{W.~S.}} \AND
\bauthor{\bsnm{Le},~\bfnm{H.}\binits{H.}}
(\byear{2009}).
\btitle{Statistical shape theory}.
In \bbooktitle{New Perspectives in Stochastic Geometry}
(\beditor{\bfnm{W.~S.}\binits{W.~S.}~\bsnm{Kendall}} \AND
\beditor{\bfnm{I.~S.}\binits{I.~S.}~\bsnm{Molchanov}}, eds.).
\bpublisher{Oxford Univ. Press}, \baddress{Oxford}.
\end{bincollection}
\MR{2654683}
\endbibitem

%
%
\bibitem[\protect\citeauthoryear{Kendall et~al.}{1999}]{kendallbook}
\begin{bbook}[author]
\bauthor{\bsnm{Kendall},~\bfnm{D.~G.}\binits{D.~G.}},
\bauthor{\bsnm{Barden},~\bfnm{D.}\binits{D.}},
\bauthor{\bsnm{Carne},~\bfnm{T.~K.}\binits{T.~K.}} \AND
\bauthor{\bsnm{Le},~\bfnm{H.}\binits{H.}}
(\byear{1999}).
\btitle{Shape and Shape Theory}.
\bpublisher{Wiley}, \baddress{Chichester}.
\end{bbook}
\MR{1891212}
\endbibitem

%
%
\bibitem[\protect\citeauthoryear{Klenow and Henningsen}{1970}]{klenow}
\begin{barticle}[author]
\bauthor{\bsnm{Klenow},~\bfnm{H.}\binits{H.}} \AND
\bauthor{\bsnm{Henningsen},~\bfnm{I.}\binits{I.}}
(\byear{1970}).
\btitle{Selective elimination of the exonuclease activity of the
deoxyribonucleic acid polymerase from escherichia coli B by limited
proteolysis}.
\bjournal{Proc. Natl. Acad. Sci. USA}
\bvolume{65}
\bpages{168--175}.
\end{barticle}
\endbibitem

%
%
\bibitem[\protect\citeauthoryear{Le and Barden}{2010}]{lebarden}
\begin{barticle}[author]
\bauthor{\bsnm{Le},~\bfnm{H.}\binits{H.}} \AND
\bauthor{\bsnm{Barden},~\bfnm{D.}\binits{D.}}
(\byear{2010}).
\btitle{On the induced distribution of the shape of the projection of a
randomly rotated configuration}.
\bjournal{Adv. in Appl. Probab.}
\bvolume{42}
\bpages{331--346}.
\end{barticle}
\MR{2675105}
\endbibitem

%
%
\bibitem[\protect\citeauthoryear{Leschziner and Nogales}{2006}]{leschziner}
\begin{barticle}[author]
\bauthor{\bsnm{Leschziner},~\bfnm{A.~E.}\binits{A.~E.}} \AND
\bauthor{\bsnm{Nogales},~\bfnm{E.}\binits{E.}}
(\byear{2006}).
\btitle{The orthogonal tilt reconstruction method: An approach to generating
single-class volumes with no missing cone for ab initio reconstruction of
asymmetric particles}.
\bjournal{J. Struct. Biol.}
\bvolume{153}
\bpages{284--299}.
\end{barticle}
\endbibitem

%
%
\bibitem[\protect\citeauthoryear{Natterer}{2001}]{naterrer}
\begin{bbook}[author]
\bauthor{\bsnm{Natterer},~\bfnm{F.}\binits{F.}}
(\byear{2001}).
\btitle{The Mathematics of Computerized Tomography}.
\bseries{Classics in Applied Mathematics}
\bvolume{32}.
\bpublisher{SIAM}, \baddress{Philadelphia, PA}.
\end{bbook}
\MR{1847845}
\endbibitem

%
%
\bibitem[\protect\citeauthoryear{O'{S}ullivan}{1995}]{osull2}
\begin{barticle}[author]
\bauthor{\bsnm{O'{S}ullivan},~\bfnm{F.}\binits{F.}}
(\byear{1995}).
\btitle{A study of least squares and maximum likelihood for image
reconstruction in positron emission tomography}.
\bjournal{Ann. Statist.}
\bvolume{23}
\bpages{1267--1300}.
\end{barticle}
\MR{1353506}
\endbibitem

%
%
\bibitem[\protect\citeauthoryear{Panaretos}{2006}]{aap}
\begin{barticle}[author]
\bauthor{\bsnm{Panaretos},~\bfnm{V.~M.}\binits{V.~M.}}
(\byear{2006}).
\btitle{The diffusion of radon shape}.
\bjournal{Adv. in Appl. Probab.}
\bvolume{38}
\bpages{320--335}.
\end{barticle}
\MR{2264947}
\endbibitem

%
%
\bibitem[\protect\citeauthoryear{Panaretos}{2008}]{mathproc}
\begin{barticle}[author]
\bauthor{\bsnm{Panaretos},~\bfnm{V.~M.}\binits{V.~M.}}
(\byear{2008}).
\btitle{Representation of radon shape diffusions via hyperspherical Brownian
motion}.
\bjournal{Math. Proc. Cambridge Philos. Soc.}
\bvolume{145}
\bpages{457--470}.
\end{barticle}
\MR{2442137}
\endbibitem

%
%
\bibitem[\protect\citeauthoryear{Panaretos}{2009}]{victorannals}
\begin{barticle}[author]
\bauthor{\bsnm{Panaretos},~\bfnm{V.~M.}\binits{V.~M.}}
(\byear{2009}).
\btitle{On random tomography with unobservable projection angles}.
\bjournal{Ann. Statist.}
\bvolume{37}
\bpages{3272--3306}.
\end{barticle}
\MR{2549560}
\endbibitem

%
%
\bibitem[\protect\citeauthoryear{Penczek, Grassucci and Frank}{1994}]{penczek}
\begin{barticle}[author]
\bauthor{\bsnm{Penczek},~\bfnm{P.}\binits{P.}},
\bauthor{\bsnm{Grassucci},~\bfnm{R.~A.}\binits{R.~A.}} \AND
\bauthor{\bsnm{Frank},~\bfnm{J.}\binits{J.}}
(\byear{1994}).
\btitle{The ribosome at improved resolution: New techniques for merging and
orientation refinement in 3D cryoelectron microscopy of biological
particles}.
\bjournal{Ultramicroscopy}
\bvolume{40}
\bpages{33--53}.
\end{barticle}
\endbibitem

%
%
\bibitem[\protect\citeauthoryear{Pisarenko}{1973}]{pisarenko}
\begin{barticle}[author]
\bauthor{\bsnm{Pisarenko},~\bfnm{V.~F.}\binits{V.~F.}}
(\byear{1973}).
\btitle{The retrieval of harmonics from a covariance function}.
\bjournal{Geophys. J. R. Astr. S.}
\bvolume{33}
\bpages{347--366}.
\end{barticle}
\endbibitem

%
%
\bibitem[\protect\citeauthoryear{{R Development Core Team}}{2011}]{R}
\begin{bbook}[author]
\bauthor{\bsnm{{R Development Core Team}}}
(\byear{2011}).
\btitle{R: A Language and Environment for Statistical Computing}.
\bpublisher{R Foundation for Statistical Computing}, \baddress{Vienna,
Austria}.
\end{bbook}
\endbibitem

%
%
\bibitem[\protect\citeauthoryear{Rademacher}{1994}]{rademacher}
\begin{barticle}[author]
\bauthor{\bsnm{Rademacher},~\bfnm{M.}\binits{M.}}
(\byear{1994}).
\btitle{Three dimensional reconstruction from random projections: orientational
alignment via Radon transforms}.
\bjournal{Ultramicroscopy}
\bvolume{17}
\bpages{117--126}.
\end{barticle}
\endbibitem

%
%
\bibitem[\protect\citeauthoryear{Sigworth}{1998}]{sigworth}
\begin{barticle}[author]
\bauthor{\bsnm{Sigworth},~\bfnm{F.~J.}\binits{F.~J.}}
(\byear{1998}).
\btitle{A maximum-likelihood approach to single-particle image refinement}.
\bjournal{J.~Struct. Biol.}
\bvolume{122}
\bpages{328--339}.
\end{barticle}
\endbibitem

%
%
\bibitem[\protect\citeauthoryear{Silverman et~al.}{1990}]{silveretal}
\begin{barticle}[author]
\bauthor{\bsnm{Silverman},~\bfnm{B.~W.}\binits{B.~W.}},
\bauthor{\bsnm{Jones},~\bfnm{M.~C.}\binits{M.~C.}},
\bauthor{\bsnm{Wilson},~\bfnm{J.~D.}\binits{J.~D.}} \AND
\bauthor{\bsnm{Nychka},~\bfnm{D.~W.}\binits{D.~W.}}
(\byear{1990}).
\btitle{A~smoothed EM approach to indirect estimation problems with particular
reference to stereology and emission tomography}.
\bjournal{J. Roy. Statist. Soc. Ser. B}
\bvolume{52}
\bpages{271--324}.
\end{barticle}
\MR{1064419}
\endbibitem

%
%
\bibitem[\protect\citeauthoryear{Tibshirani}{1996}]{lasso}
\begin{barticle}[author]
\bauthor{\bsnm{Tibshirani},~\bfnm{R.}\binits{R.}}
(\byear{1996}).
\btitle{Regression shrinkage and selection via the lasso}.
\bjournal{J. Roy. Statist. Soc. Ser. B}
\bvolume{58}
\bpages{267--288}.
\end{barticle}
\MR{1379242}
\endbibitem

%
%
\bibitem[\protect\citeauthoryear{Vardi, Shepp and Kaufman}{1985}]{vardiJASA}
\begin{barticle}[author]
\bauthor{\bsnm{Vardi},~\bfnm{Y.}\binits{Y.}},
\bauthor{\bsnm{Shepp},~\bfnm{L.~A.}\binits{L.~A.}} \AND
\bauthor{\bsnm{Kaufman},~\bfnm{L.}\binits{L.}}
(\byear{1985}).
\btitle{A statistical model for positron emission tomography (with
discussion)}.
\bjournal{J. Amer. Statist. Assoc.}
\bvolume{80}
\bpages{8--37}.
\end{barticle}
\MR{0786595}
\endbibitem

\end{thebibliography}
\end{document}